\documentclass[journal=ancham, manuscript=article, layout=twocolumn]{achemso}

\usepackage[T1]{fontenc}
\usepackage[utf8]{inputenc}
\usepackage[version=4]{mhchem} 

\usepackage{todonotes}
\usepackage{natmove}

\usepackage{algorithmicx}
\usepackage{algpseudocode}
\usepackage{algorithm}

\usepackage{amsmath}
\usepackage{graphics}
\usepackage{subcaption}
\usepackage[colorlinks=true, linkcolor=red, citecolor=red, urlcolor=red]{hyperref}
\usepackage[font=scriptsize,labelfont=bf]{caption}

\let\OLDthebibliography\thebibliography
\renewcommand\thebibliography[1]{
  \OLDthebibliography{#1}
  \setlength{\parskip}{0pt}
  \setlength{\itemsep}{0pt plus 0.3ex}
}
\algnewcommand\algorithmicinput{\textbf{INPUT:}}
\algnewcommand\INPUT{\item[\algorithmicinput]}
\algnewcommand\algorithmicoutput{\textbf{OUTPUT:}}
\algnewcommand\OUTPUT{\item[\algorithmicoutput]}

\author{Mateusz K. Łącki}
\email{mateusz.lacki@biol.uw.edu.pl}
\affiliation{Department of Mathematics, Informatics, and Mechanics, University of Warsaw, 02-097 Warsaw, Poland}

\author{Frederik Lermyte}
\affiliation{Biomolecular and Analytical Mass Spectrometry Group, Department of Chemistry, University of Antwerp, Antwerp, Belgium}
\alsoaffiliation{Centre for Proteomics, University of Antwerp, 2000 Antwerp, Belgium}

\author{Błażej Miasojedow}
\affiliation{Department of Mathematics, Informatics, and Mechanics, University of Warsaw, 02-097 Warsaw, Poland}

\author{Michał P. Startek }
\affiliation{Department of Mathematics, Informatics, and Mechanics, University of Warsaw, 02-097 Warsaw, Poland}

\author{Frank Sobott}
\affiliation{Biomolecular and Analytical Mass Spectrometry Group, Department of Chemistry, University of Antwerp, Antwerp, Belgium}
\alsoaffiliation{Astbury Centre for Structural Molecular Biology, University of Leeds, Leeds, UK}
\alsoaffiliation{School of Molecular and Cellular Biology, University of Leeds, Leeds, UK}

\author{Dirk Valkenborg}
\affiliation{Centre for Proteomics, University of Antwerp, 2000 Antwerp, Belgium}
\alsoaffiliation{Flemish Institute for Technological Research (VITO), 2400 Mol, Belgium}
\alsoaffiliation{Interuniversity Institute for Biostatistics and Statistical Bioinformatics, Hasselt University, 3500 Hasselt, Belgium}

\author{Anna Gambin}
\affiliation{Department of Mathematics, Informatics, and Mechanics, University of Warsaw, 02-097 Warsaw, Poland}

\title[MassTodon]{Assigning peaks and modeling ETD in top-down mass spectrometry}


\begin{document}

\begin{abstract}
Among many techniques of modern mass spectrometry, the top down methods are becoming continuously more popular in the strife to describe the proteome. 
These techniques are based on fragmentation of ions inside mass spectrometers instead of a  proteolytic digestion. 
In some of these techniques, the fragmentation is induced by electron transfer.
It can trigger several concurrent reactions: electron transfer dissociation, electron transfer without dissociation, and proton transfer reaction.  
The evaluation of the extent of these reactions is important for the proper understanding of the functioning of the instrument. 
It is even more important, to know if it can be used to reveal important structural information.

We present a workflow for assigning peaks and interpreting the results of electron transfer driven reactions.
We also present software code-named {\tt MassTodonPy} available for use free of charge under the GNU GPL v3 license.
\end{abstract}

\section{Introduction}
In recent years, there has been growing interest in electron-based dissociation (ExD) – primarily electron capture (ECD) \cite{zubarev1998electron} and electron transfer dissociation (ETD)\cite{Syka2004PeptideAndProtein} in protein mass spectrometry.
These fragmentation methods allow the cleavage of the backbone of a protein or peptide without significantly disrupting other bonds (even preserving noncovalent interactions) and as such, much effort has gone into the use of ExD methods for top-down sequencing, as well as the study of labile post-translational modifications and even binding sites of non-covalent ligands\cite{garcia2007characterization,haakansson2001electron,ayaz2009vivo,ge2009top,tsybin2011structural,fornelli2012analysis,cournoyer2005deamidation,li2010glutamine,xie2006top,jackson2009use,yin2010elucidating,goth2016gas}. 
Additionally, considerable efforts have been made to determine preferential reaction pathways and cleavage sites in ExD of known precursors, to obtain insight into gas-phase protein/peptide conformation \cite{breuker2002detailed,oh2002secondary,skinner2012ubiquitin,skinner2013charge,zhang2011native,zhang2013native,zhang2014exploring,lermyte2014etd,lermyte2015electron,zhang2016native,lermyte2017conformational} as well as to investigate the reaction mechanism \cite{turecek2003n,turecek2003mechanism,chung2010backbone}. 
Ideally, reaction products are not only identified, but also quantified in these efforts. Because of the information-rich nature of top-down ExD spectra, data processing is usually performed with the help of specialized software.

The first, and arguably most critical step in this data processing is usually spectral deisotopisation, i.e. reducing the multitude of signals observed in the m/z dimension due to various charge states and isotopologues to a minimal set of components and abundances.
Most of the readily available software tools for this – e.g. {\tt THRASH}\cite{horn2000automated}, {\tt MASH}\cite{guner2014mash,cai2016mash}, {\tt DeconMSn}\cite{mayampurath2008deconmsn}, {\tt Decon2LS}\cite{jaitly2009decon2ls} – utilize an averagine-scaling approach\cite{senko1995determination} to determine charge states, monoisotopic masses, and ion intensities.
As this requires resolution of the (aggregated) isotope peaks, these tools are mostly used to process FTICR or Orbitrap data, particularly as they can natively process Bruker and/or Thermo data files (in fact, a modified {\tt THRASH} algorithm, called {\tt SNAP}, is built into the Bruker DataAnalysis software). 

Observed isotope clusters are often composed of multiple overlapping isotope distributions (envelopes), each  generated by ions whose chemical formulas differ by one (or a few) hydrogen atoms. 
These shifts (by an integer number of hydrogen masses) are commonly observed in ExD spectra and provide information on reaction pathways\cite{lermyte2017conformational,o2006long,tsybin2007ion}. 
As such, it is desirable to preserve the information contained in observed isotope distributions during and after the deconvolution procedure.

Thus, there is a need for software tools which are able to process high-resolution tandem MS data from a variety of instruments, utilize the high-resolution information (e.g. properly assign highly resolved peaks) to perform thorough data analysis, and provide the user with information regarding preferred cleavage sites and relative probabilities of competing reaction pathways.
Ideally, this should not require the user to possess extensive expertise regarding statistics and/or gas-phase ion/ion chemistry.
Recently, we have demonstrated the use of an in-house developed software for deconvoluting complex isotope clusters occurring in top-down ETD spectra acquired on a Waters Synapt G2 Q-IM-TOF instrument\cite{lermyte2015understanding}.
Furthermore, we have shown how this allows us to infer branching ratios and how this correlates to collision cross-sections and gas-phase conformations of ubiquitin\cite{lermyte2017conformational}.
Here, we present in detail the above computational workflow, together with extensions that shed further light onto the electron transfer driven reactions.
The Python implementation of that workflow, called {\tt MassTodonPy}, is made publicly available for download via Python Package Index. 

\noindent\textbf{Paper organization.} 
In the rest of the article we describe the stages of the proposed workflow: (1) the preprocessing of the spectrum, (2) the generation of potentially observable chemical formulas, (3) the deconvolution of spectra, which involves the estimation of the intensities of the potential products of the considered set of reactions, (4) the pairing of fragment ions, resulting in estimates of the probabilities of the considered reactions and fragmentations.
The workflow was tested \textit{in silico} and on around 200 mass spectra. 
Finally, we mention some possible extensions to the workflow.

\begin{table*}[t]
\centering
\begin{tabular}{rlcl}
 	\textbf{PTR} 	&\ce{[M + nH]^{n+} + A^{.-}} 	&\ce{->}& \ce{[M + (n-1) H]^{(n-1)+} + AH}  \\
 	\textbf{ETnoD} 	&\ce{[M + nH]^{n+} + A^{.-}} 	&\ce{->}& \ce{[M + nH]^{(n-1)+.} + A}      \\
 	\textbf{ETD} 	&\ce{[M + nH]^{n+} + A^{.-}} 	&\ce{->}& \ce{[C + xH]^{x+} + [Z + (n - x)H]^{(n-x-1)+.} + A}\\
 	\textbf{HTR} 	&\ce{[C + xH]^{x+}} 			&\ce{->}& \ce{[C + (x - 1)H]^{x+}}\\
 	 			 	&\ce{[Z + (n - x)H]^{(n-x-1)+}} 	&\ce{->}& \ce{[Z + (n - x + 1)H]^{(n-x-1)+}}
\end{tabular}
\caption{Considered chemical reactions. \ce{M} stands for either a precursor ion or a fragment ion. The HTR reaction can happen only after ETD and consists in the transfer of a hydrogen atom from the $c$ to the $z$ fragment.}\label{chemical_reactions}
\end{table*}

\section{Materials and methods}
\noindent\textbf{Data Preprocessing.} 
We assume that the input spectrum was already calibrated. 
The spectrum should not be centroided, as {\tt MassTodon} does its own centroiding, as described later in the peak picking section.

To attenuate the possibility of fitting to noise peaks, some parts of the mass spectrum need to be trimmed out. 
We offer two simple ways to do it.
The first way focuses on the intensity of individual peaks and amounts to trimming out peaks with intensity below a user-provided threshold.
The second way retains only the heighest peaks whose joint intensity cover the user specified percentage of the total intensity in the spectrum.
To make that idea more clear, consider a spectrum comprised of three peaks with intensities equal to 1000, 990, and 10. 
Also, set the joint threshold at 99\%. 
The intensity of the first peak amounts to $\frac{1000}{1000+990+10} = 50\%$ of the entire intensity in the spectrum. 
The intensity of the first two peaks amounts to $\frac{1000+990}{1000+990+10} = 99.5\%$ of the overal intensity. It is the smallest set of heighest peak that jointly surpass the required threshold of $99\%$ and so only these peaks are left, and the third one is trimmed out. 
Observe that the same effect would be achieved if trimming out peaks with intensity above 990. 
For each run of the second trimming spectrum we calculate that implicit cut-off and store it for the inspection of the user.

Finally, the mass to charge ratios are rounded to better match the theoretical spectra, as described later on.

\noindent\textbf{Generating chemical formulas.} 
{\tt MassTodon} exhausetively finds the formulas of all molecular species that might be present in the set of considered reactions.
The theoretical envelopes of these molecules are then fitted to the spectral data in a later stage. 

The presented workflow considers a set of known chemical reactions triggered by the electron transfer, c.f. Table~\ref{chemical_reactions}. 
The Proton Transfer Reaction (PTR) and the non-dissociative Electron Transfer Dissociation (ETnoD) do not result in any fragments; they affect the charge state and the mass of the cation alone. 
The Electron Transfer Dissociation (ETD), potentially followed by the transfer of a hydrogen (HTR), result in $c$ and $z$ fragments\citep{RoepstorffScheme}. 
We assume that PTR and ETnoD may occur multiple times on the same ions, including the $c$ and $z$ fragments.
We assume that fragments cannot further fragment, as the inner fragments are scarcely ever observed experimentally in ETD.
The number of fragments depends on the charge of the precursor filtered during $\text{MS}_1$, denoted $Q$, its amino acid sequence and the existing modifications. 
We neglect the ordering of reactions within one pathway.
Thus, the product of the PTR reaction followed by the ETnoD reaction is the same as the product of the ETnoD reaction followed by the PTR reaction. 
In general, reaction pathways leading to the same product are indiscernible until the last stage of the algorithm. 

Every molecular species is described by its elemental composition and charge $q$. 
Each reaction consumes one charge.
During PTR, the radical passes from anion to cation reducing its charge without significantly changing its mass (we neglect the mass of the electron). 
This motivates the introduction of an additional quantity, the \textit{quenched charge} $g$, that describes the number of extra hydrogen masses with respect to precursor's hydrogen content, see \citet{lermyte2015understanding}.
An increase in $g$ corresponds to an increase in one atomic mass unit and does not change the charge state. 

To exemplify the above concept, consider triply charged Substance P, \ce{\text{\tt RPKPQQFFGLM}^{3+}}. 
The mass of its monoisotopic isotopologue equals $1347.712$[u], when rounded to the third decimal place.
Add the mass of two protons and one quenched charge and divide it by the two present charges to get $\frac{1347.712 + 3 \times 1.008}{2} = 675.368$[Da].
Thus, the regions of the mass spectrum close to that value can contain ions belonging to that molecular species.
Consider the case of filtering only triply charged precursors during the MS1, i.e. selecting ions with m/z around $450.245$[Da]. 
It is then possible to state that these ions must have underwent exactly one ETnoD.
This is because ETnoD would reduce their charge by one, \ce{3^+ -> 2^+}, and increases the number of quenched charges by one, see Table~\ref{chemical_reactions}.
Further on we show how to infer the number of reactions both from precursor ions and fragments.

While studying the above example, it is important to notice that other sources of ions can explain the same peak.
In particular, consider the second most probable isotopologue of the precusor ion that underwent the PTR reaction.
One of the \ce{^{12}C} carbon atoms in this isotopologues is exchanged for a heavier isotopic variant, \ce{^{13}C}. 
These ions are only slightly less likely to be found in the sample than the monoisotopic ions: on average in $29.6\%$ of cases for this ion source versus $43.1\%$ for the monoisotopic peak.
Their mass is $1348.716$[u]. When equiped with two charges, their m/z equals $\frac{1348.716 + 2 \times 1.008}{2} = 675.366$.
Most instruments would not resolve the $0.002$[Da] difference between the two molecular species.  
However, confusing the two ions sources leads to a poor estimate of the relative relative extent of PTR versus ETnoD.  
Based on one peak alone it is impossible to correctly identify the relative proportions of different molecular species.
However, in most cases it is possible to differentiate between various molecular species by looking at their isotopic distributions as a whole.
This opens the possibility to evaluate how much of observed intensity can be attributed to particular ion sources.
Further on we show how this can be achieved.

Observe that quenched charge may also be used to record information on a hydrogen transfered during HTR.
This is convenient, as there is nor real difference between a quenched charge and a \textit{regular} hydrogen atom within one molecule.
Consider then a precursor that undergoes a direct HTR reaction: the c fragment must then have a quenched charge equal to -1, which we consider a valid possibility. 
In that case alone does this quantity assume a negative value.

During the fragmentation, the remaining charge and quenched charge (if positive) are distributed among the fragments. 
One might expect the charge state of smaller fragments to be limited, due to Coulomb repulsion.
For this reason, \textsc{MassTodon} omits formulas with too many charges per a given number of amino acids, the default being set to 5. 
In case of Substance P, this means that we could not observe a $c_{3}$ fragment with two charges, but we could observe a $c_5$ fragment.
The charge distance parameter can be adjusted by the user.

If one considered only the PTR and ETnoD reactions, the precursor molecule could result exactly in $\frac{Q(Q+1)}{2}$ different molecular species. 
Each product can be further fragmented into pairs of different $c$ and $z$ fragments. 
The number of such pairs is $K$ - the number of amino acids in the provided sequence, minus the number of prolines, that cannot be fragmented easily by electron transfer due to their ring structure. 
Then, each fragment can again undergo several PTR and ETnoD reactions. 
The number of all fragments is thus of the order of $\mathcal{O}(K Q^4)$.

\begin{figure*}[t]
	\begin{subfigure}[b]{0.7\textwidth}
		\centering
		\includegraphics[width=\linewidth]{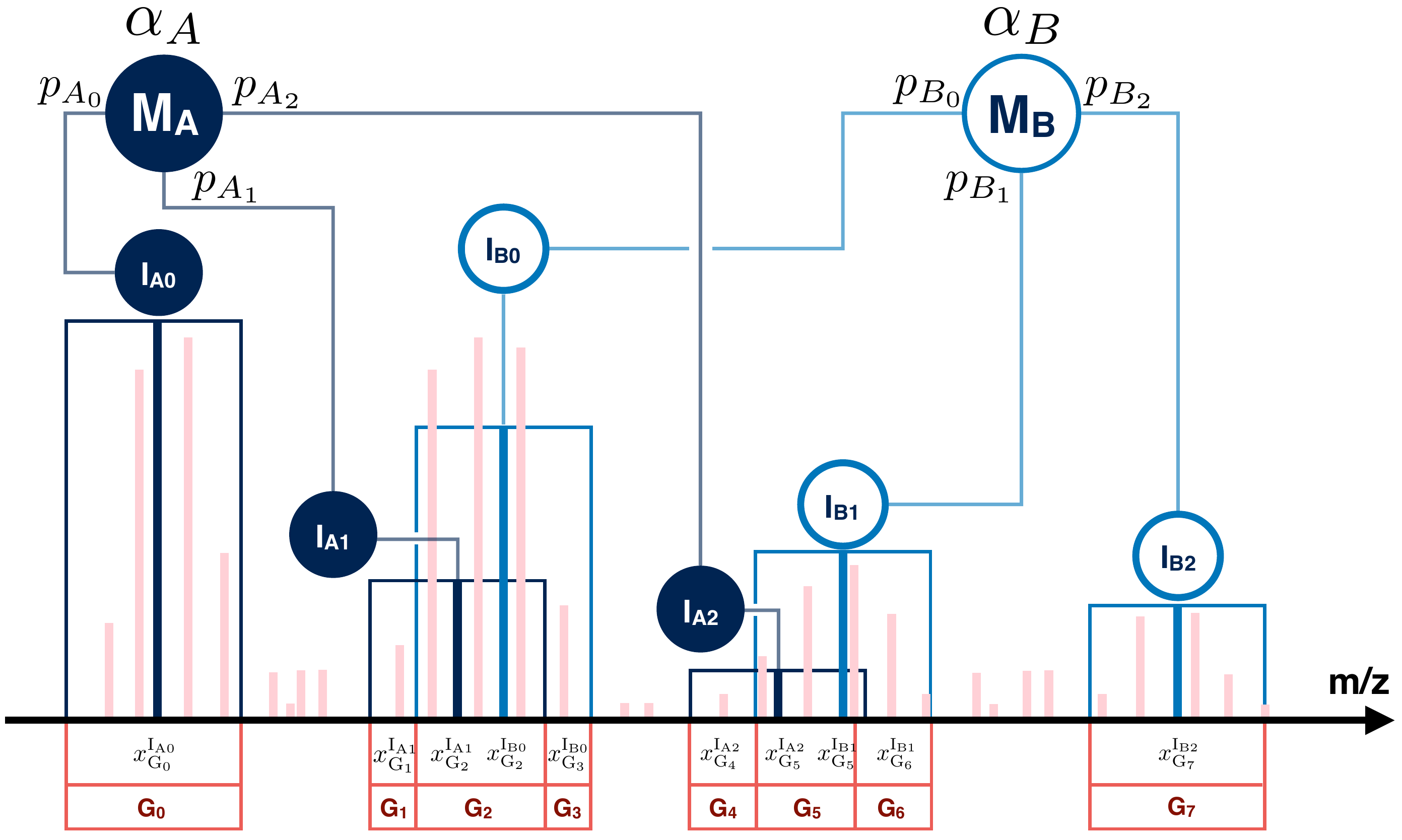}
		\caption{$\mathcal{C}$, a connected component of $\mathcal{G}$}\label{fig::deconvolution_principles}
	\end{subfigure}
	\begin{subfigure}[b]{0.29\textwidth}
		\begin{subfigure}[b]{\textwidth}
	    \includegraphics[width=\textwidth]{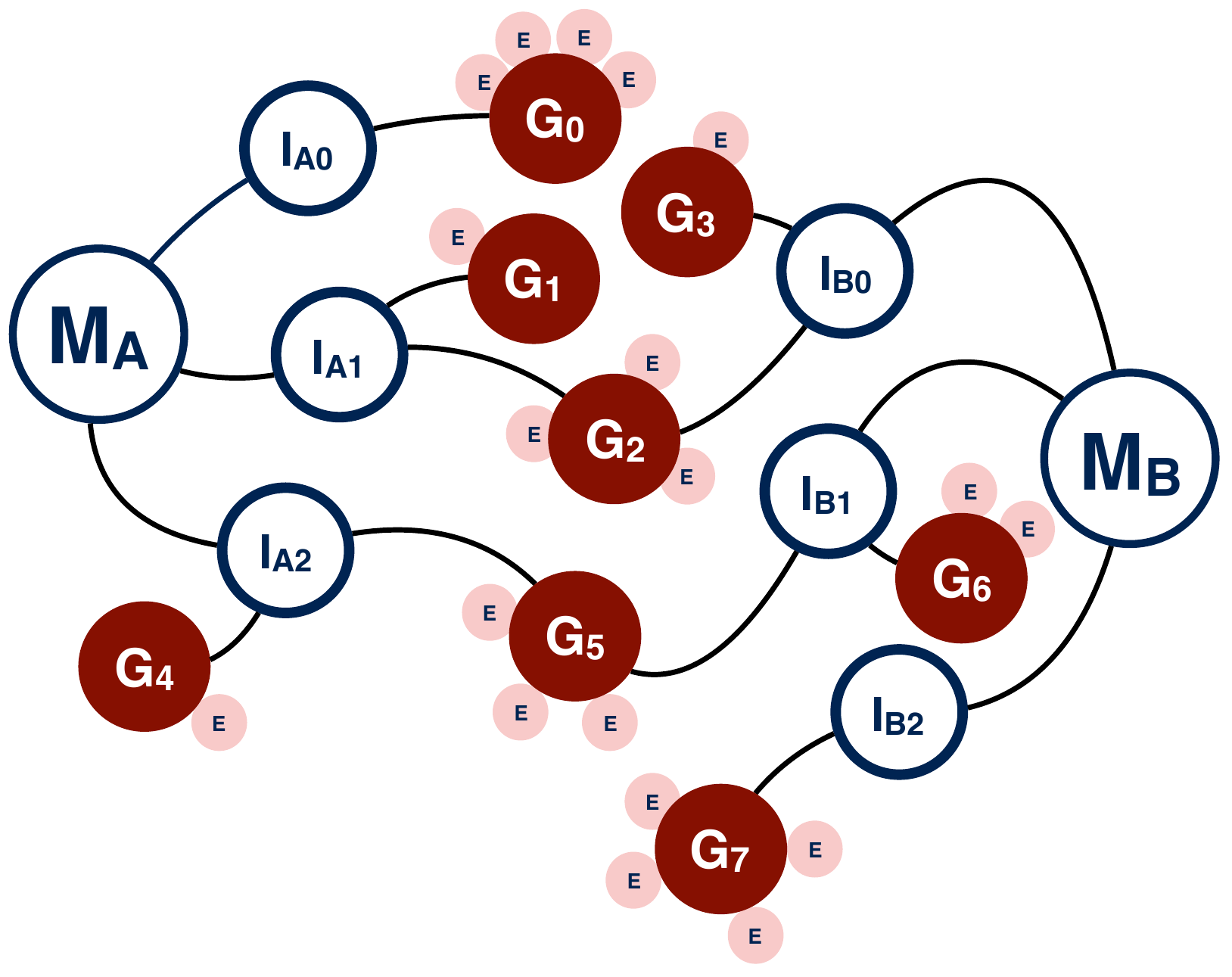}
	    \caption{Graph form of $\mathcal{C}$}
	    \label{fig::graph form of G}
		\end{subfigure}
		~ 
		\begin{subfigure}[b]{\textwidth}
			\centering
		    \includegraphics[width=.6\textwidth]{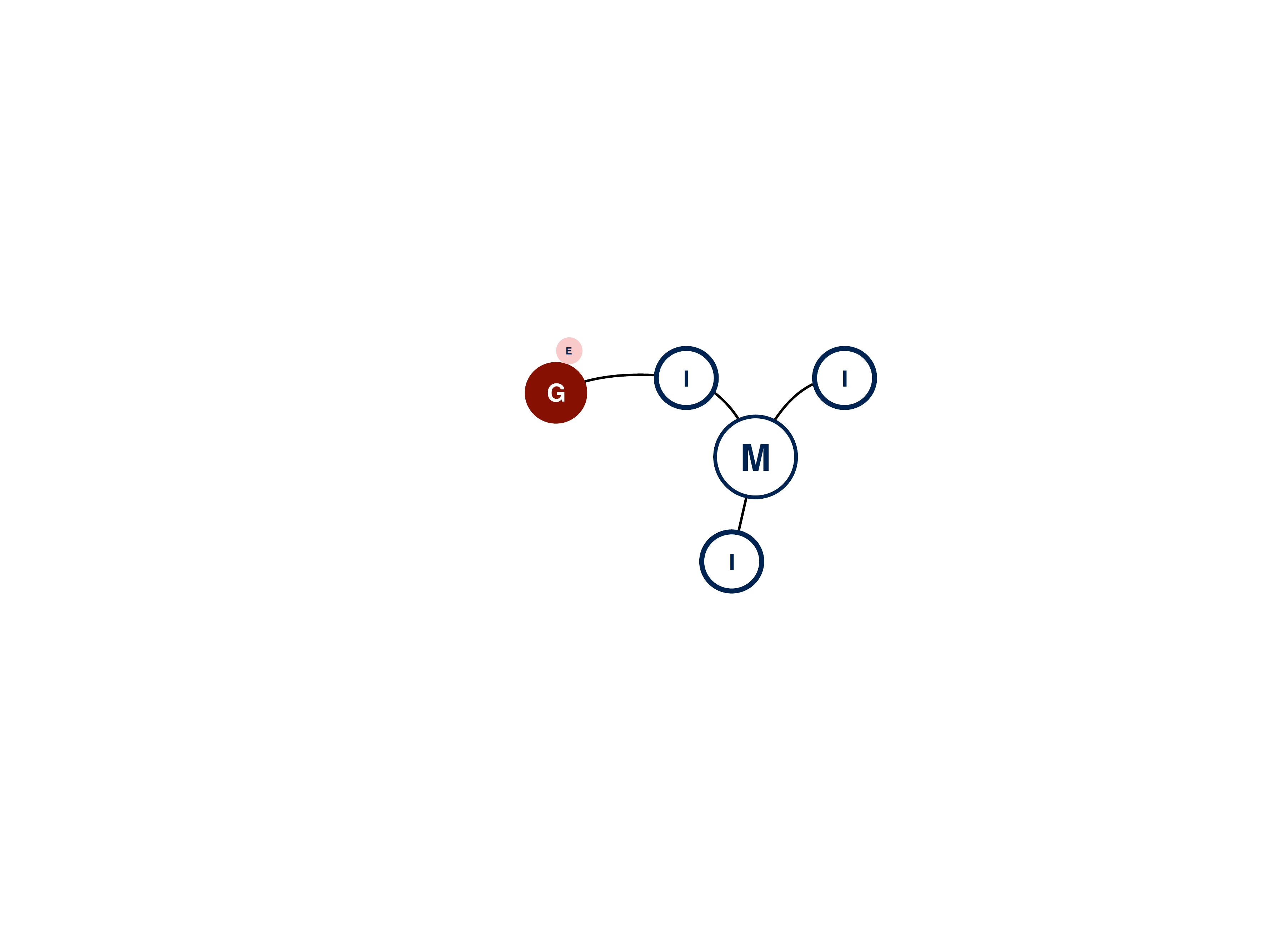}
		    \caption{A molecule with with scarse experimental support}
		    \label{fig::poor support}
		\end{subfigure}
	\end{subfigure}
	\caption{A connected component $\mathcal{C}$ of the \textit{deconvolution graph} $\mathcal{G}$. 
    Experimental peaks are shown in pink.
	Among the nodes of $\mathcal{G}$ we find the molecules $M$, their isotopologues $I$, and experimental groups $G$. 
    The probability $p$ of meeting $I$ among the $M$ ions decorates the edge between $I$ and $M$. 
    Edges between $I$ and $G$ are not plotted for clarity in (a); we do mark however their corresponding flow variables, $x$. 
    They denote the amount of experimental intensity attributed to a given isotopologue.
	The aim of the deconvolution is to establish total intensities of $\text{M}_1$ and $\text{M}_2$, denoted respectively as $\alpha_A$ and $\alpha_B$. In (b) we show $\mathcal{C}$ as a graph. 
    The experimental peaks (in pink) are depicted only for clarity of the representation and are not actually in $\mathcal{G}$.
    In (c) we show a molecule $M$ with scarse experimental support.}
\end{figure*}

\noindent\textbf{Generating the isotopic distributions.}
The isotopic distribution of a given molecular species models the expected signal one could register in the mass spectrometer. 
In terms of the presented workflow, it offers a way to relate the intensities of peaks in the mass spectrum with different m/z ratios, whenever they follow a predicted pattern.

Each reaction product is described by its elemental composition, charge $q$, and quenched charge $g$. 
This information is sufficient to generate the theoretic isotopic distibution using any isotopic calculator. 
To perform calculations here, we use the \textsc{IsoSpec} algorithm \cite{lacki2017isospec}. 
Given the elemental composition, \textsc{IsoSpec} produces a series of infinitely resolved isotopologues, representable as tuples (mass, probability). 
To avoid the combinatorial explosion in their number\cite{Valkenborg2012Isotopic}, \textsc{IsoSpec} reports only the smallest possible set of peaks, such that their cumulative probability does not fall under some user specified threshold, e.g. $99.9\%$. 
The masses of the envelopes are adjusted according to formula $\frac{m+q+g}{q}$ to obtain valid mass over charge ratios.

To model low resolution spectra, one does not need infinitely resolved theoretical envelopes. 
Whenever small differences between the m/z ratios cannot be discerned, one can safely aggregate peaks with similar m/z ratios.
In our workflow, we ask the user to provide a measure of the instrument's resolution in terms of one parameter alone -- the peak's m/z tolerance $tol$.
Experimental peaks are deemed to potentially originate from a molecule $M$ if their m/z ratios are within the $tol$ distance from a theoretical isotopologue $I$ of that molecule. 
This is shown in Figure~\ref{fig::deconvolution_principles}. 
By default, we assume that differences between m/z ratios an order of magnitude smaller than $tol$ cannot be discerned. 
This implies a finite granularity of the spectrum: if $tol$ amounted to 0.05 [u], then the smallest difference between peaks would be that of 0.001 [u].  
To obtain such spectrum, peaks with the same first three significant digits are aggregated, i.e. they are represented by one peak with the same rounded m/z and intensity equal to the total intensity of these peaks. 
In general, given tolerance $tol$, we round the spectrum to the significant digit given by $\lceil - log_{10}(tol)\rceil$ and then aggregate it. 
By convention, we still call the so obtained cluster of isotopologues an isotopologue.
The same operation is performed on the experimental m/z ratios.
This step reduces the size of the deconvolution problem and speeds up the peak picking and the deconvolution.
That said, one should not provide an underestimate of $tol$ to speed up computations.
This is because highly resolved spectra offer the possibility to discern between isotopologues of different molecular species and to better identify their joint intensities.

\noindent\textbf{Peak picking.}
The aim of the peak picking is to assign peaks in the mass spectrum to the potential molecular species. 
This is done by comparing the m/z ratios of the experimental peaks with these of the peaks in the theoretical isotopic envelopes, as described in the previous section and visualized in Figure~\ref{fig::deconvolution_principles}. 
Figure~\ref{fig::deconvolution_principles} also shows that finding potential explanations for a given experimental peaks corresponds to finding all intervals of the form $[\frac{m}{z} - tol, \frac{m}{z} + tol]$ to which its m/z value belongs. 
To find these intervals effectively, we make use of the interval trees data structure \citep{cormen2009leiserson}.

Different intervals might overlap, as is the case for isotopologues $I_{A1}$ and $I_{B0}$ in Figure~\ref{fig::deconvolution_principles}.
The intersections of these intervals partition the m/z axis into regions that can be traced back to originate from different sets of molecules and regions that cannot be explained by any of the products of the considered reactions. 
Experimental peaks inside such intersections (there might be more then one) form experimental groupings $G$.
The total intensity within one such groupings is stored and denoted by $G_\text{\tt intensity}$. 
After these operations, the experimental peaks do not play any more role in calculations and can be deleted.

Considered together, molecules $M$, their isotopologues $I$, and the experimental groupings $G$ form nodes of the \textit{deconvolution graph}, $\mathcal{G}$, as shown in Figures~\ref{fig::deconvolution_principles} and \ref{fig::graph form of G}.
In $\mathcal{G}$, molecule nodes $M$ are naturally joined with their isotopologue nodes $I$, that are in their turn joined with experimental groupings $G$ they could explain.
Graph $\mathcal{G}$ is usually composed of several connected components, like the one presented in Figure~\ref{fig::deconvolution_principles}.

While picking the peaks, one can easily spot molecules $M$ with poor experimental support, as shown in Figure~\ref{fig::poor support}.
More precisely, if the sum of probabilities of isotopologues of $M$ connected to some $G$ does not exceed some percentual threshold $P$ (by default, $70\%$), then we can discard it.  
This additional preprocessing eliminates substances that alone could not explain more than the $P$ percent of the total experimental intensity within the considered subproblem, and thus makes part of the overall variable selection procedure we consider.

Each connected component of $\mathcal{G}$ gives rise to some deconvolution problem, as several molecules might compete for the explanation of the given range of the mass spectrum. 
These problems might be solved independently and simultaneously rather than sequentially. 
{\tt MassTodonPy} offers both ways of performing these calculations.

\noindent\textbf{Deconvolution.} 
The problem of deconvolving the intensities within one connected component of graph $\mathcal{G}$ is reminiscent of linear regression.
Indeed, the goal is to express the observed signal as a weighted sum of the isotopic envelopes. 
Denoted the weight by $\alpha$, as in Figure~\ref{fig::deconvolution_principles}. 
It can be interpreted as the total intensity of a given molecular species in the mass spectrum. 
In particular, $\alpha$ cannot be negative.

In advance, one does not know how to redistribute the intensity of $I$ among the neighboring experimental groupings $G$. 
This motivates the introduction of the \textit{flows} between $G$ and $I$, denoted by $x_G^I$, 
For instance, in Figure~\ref{fig::deconvolution_principles} isotopologue $I_{B0}$ is linked with experimental intentensities $G_2$ and $G_3$. 
It absorbs $x_{G_2}^{I_{B0}}$ of the intensity of $G_2$, and $x_{G_3}^{I_{B0}}$ of the intensity of $G_3$. 
$I_{B0}$ should contribute $x_{G_2}^{I_{B0}} + x_{G_3}^{I_{B0}}$ to $M_B$.
On the other hand, this should be equal to a fraction $p_{B_0}$ of the total intensity of $M_B$, denoted by $\alpha_B$. 
In other words, $p_{B_0} \alpha_B = x_{G_2}^{I_{B0}} + x_{G_3}^{I_{B0}}$.
Similarly, in general the intensities of isotopologues $I$ and molecules $M$ are related via a set of linear restrictions $\alpha_M p_M^I = \sum_{G: G \leftrightarrow I} x^I_G$, where under the sum we iterate over all experimental groups $G$ that neighbor isotopologue $I$. 

It is sensible to choose molecular intensities $\alpha$ and isotopologue intensities $x$ to assure a minimial divergence between the observed group intensities $G_\text{\tt intensity}$ and the total outflows of intensity from these nodes towards the isotopologue nodes.
The overall deconvolution problem can thus be formalized as 
\begin{align*}
	\min_{x,\alpha} \sum_{G} \big(G_\text{intensity} - \sum_{I: G\leftrightarrow I} x^I_G \big)^2\quad\text{so that}\\
    \alpha_M p_M^I = \sum_{G: G \leftrightarrow I} x^I_G, \quad x^I_G \geq 0
\end{align*}
To minimize the risk of numerical instability and perform model selection one can include in the cost function additional penalty terms\cite{james2013introduction},
{\footnotesize\begin{align*}
     L_1^x \sum_{G\leftrightarrow I} x^I_G + L_1^\alpha \sum_M \alpha_M + L_2^x \sum_{G\leftrightarrow I} (x^I_G)^2 + L_2^\alpha \sum_M \alpha_M^2.
 \end{align*}}  
By default, we set $L_1^\alpha, L_2^\alpha, L_1^x,$ and $L_2^x$ to $0.001$.
The penalty terms after $L_1^\alpha$ and $L_1^x$  should round small estimates to zero, as in the lasso model selection approach\cite{james2013introduction}.
The above problem can be efficiently solved with quadratic programming. 
MassTodon relies on the {\tt CVXOPT} Python module\cite{andersen2013cvxopt} that
solves quadratic programmes with a path following algorithm. 

After each deconvolution, we calculate and report various error statistics. 
These include the sum of the absolute values of the errors, the sum of overestimated values, and the sum of the underestimated values. 
The above quantities are also divided by the total ion current or the total intensity within the tolerance regions of any of the theoretically molecular species.

The cost function is minimized simultaneously in $x$s and $\alpha$s. 
Only $\alpha$s are analyzed in the next, final stage of the algorithm.

\noindent\textbf{Pairing of the observed ions.}
Up to this step, the algorithm obtained estimates of intensities of each considered product molecule, uniquely defined by its type (precursor, $c$ or $z$ fragment), charge $q$, quenched charge $g$.
Previously\cite{lermyte2017conformational}, we described a method for the retrieval of information on the branching ratios, i.e. the probabilities of ETnoD and PTR, entirely based upon estimates of the intensities of the non-fragmented ions. 
Given a non-fragmented molecular species with charge $q$ and quenched charge $g$,  one can retrieve the numbers of the PTR and ETnoD reactions by solving
\begin{align}
	q &= Q - N_\text{PTR} - N_\text{ETnoD},\label{eq::q} \\ 
	g &= N_\text{ETnoD},\label{eq::g}
\end{align}
for $N_\text{PTR}$ and $N_\text{ETnoD}$.
Eq.~\eqref{eq::q} states that each reaction reduces the observed charge by one. 
Eq.~\eqref{eq::g} traces the origin of all quenched charges on the precursor molecules solely to the ETnoD reaction. 
The estimate of the probability of ETnoD then equals
\begin{equation*}
	\hat{p} = \frac{ \sum_i N^i_\text{ETnoD}I_i }{ \sum_i (N^i_\text{ETnoD} + N^i_\text{PTR}) I_i }.
\end{equation*}
The index $i$ iterates over different observed precursors.
The nominator counts ions that underwent ETnoD, $I_i$ is the estimated intensity of the precursor with charges $(q_i, g_i)$. 
The denominator additionally contains the count of ions undergoing PTR.
The above estimator relies on the presumed proportionality of the signal intensity to the actual number of molecules. 

With the above method one cannot retrieve the probabilities of fragmentations.
This is because counts of reactions are not directly accessible and only estimates of the overall intensity of $c$ and $z$ fragments are at hand. 
To unveil the number of fragmentation events, one has to pair back the matching $c$ and $z$ fragments.

\begin{figure}[t]
    \begin{subfigure}[b]{0.55\linewidth}
        \includegraphics[width=\textwidth]{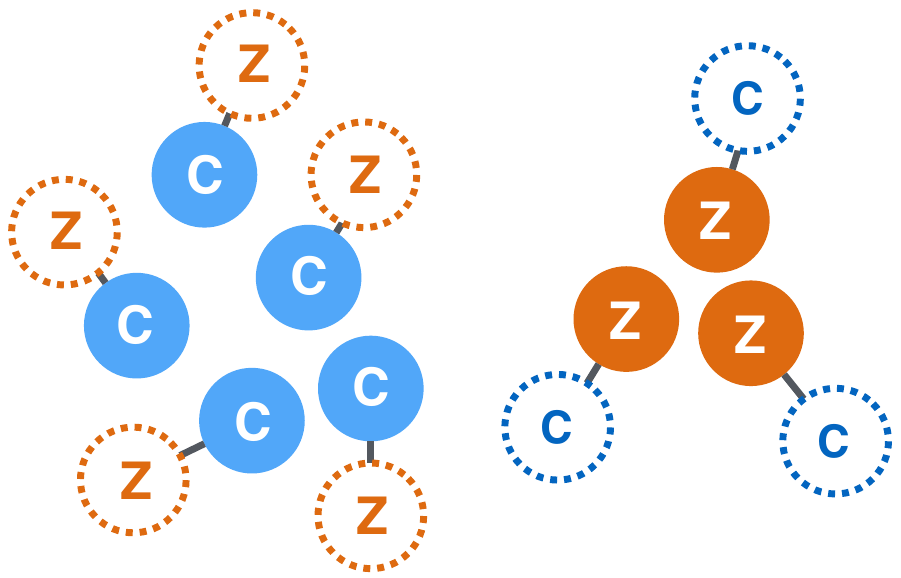}
        \caption{Lavish}
        \label{fig::lavish pairing}
    \end{subfigure}
    \begin{subfigure}[b]{0.4\linewidth}
    	\centering
        \includegraphics[width=.7\textwidth]{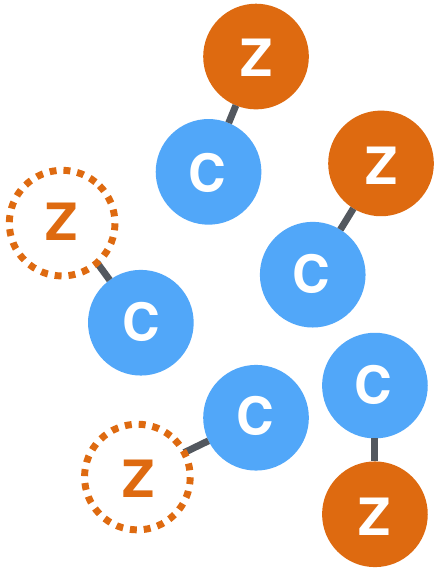}
        \caption{Parsimonious}
        \label{fig::parsimonious pairing}
    \end{subfigure}
    \caption{Two interpretations of observing 5 $c$ and 3 $z$ matching fragments: lavish (a) and parsimonious (b). Nodes with dashed edges symbolize cations that never reach the detector. (a) maximizes the number of missing cations needed to explain the spectrum, while (b) minimizes that number.
    }\label{fig::pairing problem}
\end{figure}

There exists a whole range of possible pairing strategies. 
The two extremes are: (1) ions come from entirely separate groups of precursors, and (2) the observed fragments are generated by a minimal number of precursors. 
For instance, in Figure~\ref{fig::pairing problem} we show a situation where 5 $c$ and 3 $z$ matching fragments were observed (filled circles).
In principle, one could say, that at the beginning of the experiment there were together 8 precursor molecules, but after the fragmentation one of each fragments always lost all of its charge.
This is the \textit{lavish} interpretation, as shown in Figure~\ref{fig::lavish pairing}.
If the ions could not be observed only because of loosing the entire charge, then this scenario would require a lot of reaction events to explain the outcome.
Figure~\ref{fig::parsimonious pairing} depicts the other possibility. 
Here, a maximal pairing is performed, and only two $c$ fragments have to be paired with $z$ fragments with a depleted charge (dashed circles). 
This approach is much more \textit{parsimonious} in terms of reactions needed to explain the experimental results.

Irrespectful of the above strategies, only matching ions should be paired, i.e. a $c_k$ fragment should be matched only with a $z_{K-k}$ fragment, where $K$ is the total number of amino acids in a given sequence. 
Moreover, pairing should include natural restrictions on the charge states $(q_c,q_z)$ and quenched charges $(g_c,g_z)$ of both fragments. 

\begin{figure*}[t]
	\begin{subfigure}[b]{0.49\linewidth}
		\centering
		\includegraphics[width=.7\linewidth]{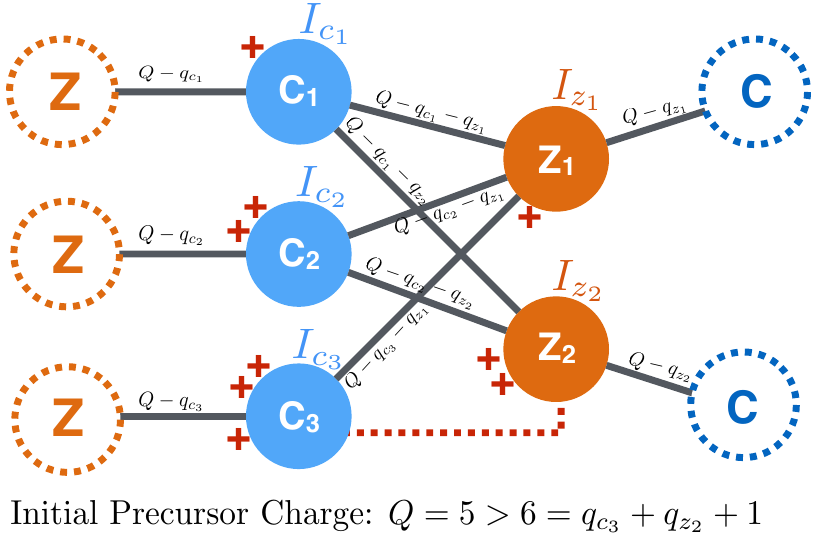}
		\caption{Pairing Graph}
        \label{fig::pairing graph}
	\end{subfigure}
	\begin{subfigure}[b]{0.49\linewidth}
		\centering
		\includegraphics[width=\linewidth]{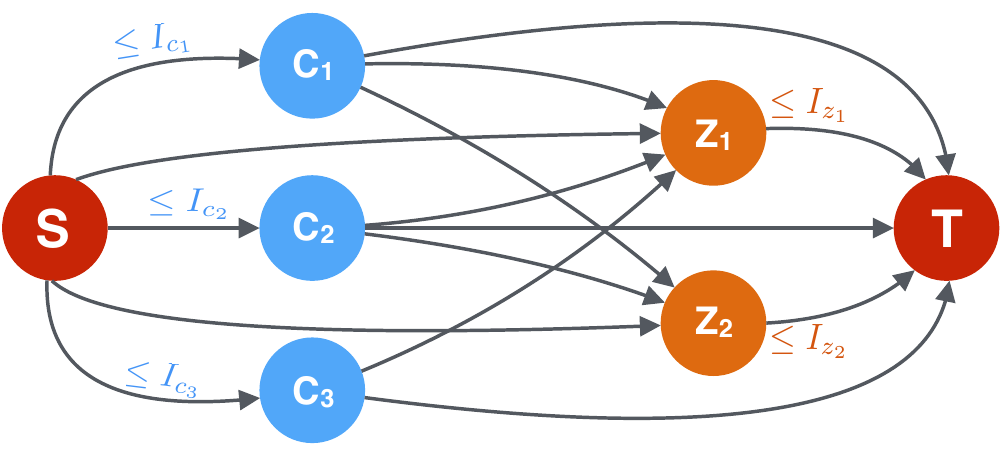}
		\caption{Max Flow Problem}
        \label{fig::max flow problem}
	\end{subfigure}
	\caption{ A \textit{pairing graph} (a) and it rrepresentation as a \textit{max flow} optimization problem (b). Nodes with dashed edges correspond to ions that lost their charge; other nodes correspond to observed fragments. In (a), ion charge are shown as red plus signs. Gray edges mark possible pairings. Red dashed line between $c_3$ and $z_2$ marks an impossible pairing: if combined, both fragments must have originated from a 6+ precursor, which was not possible. The task is to redistribute the intensity in nodes along the edges. This comes at a cost $Q - q_c - q_z$. To turn (a) into (b), one has to: (1) remove unobserved ion nodes (2) direct remaining edges from $c$ to $z$ (3) add sink S and terminal T (4) add edges directed from S to $c$ nodes and from $z$ nodes to T and add capacities equal to observed ion intensities (5) add edges from S to $z$ fragments and edges from $c$ fragments to T: these correspond to pairings with unobserved ions. This representation is possible for \textit{basic} and \textit{intermediate} pairing algorithms.
	}\label{fig::pairing to max flow}
\end{figure*}

The \textit{basic algorithm} we propose to solve the pairing problem disregards quenched charges $g_c$ and $g_z$: intensities of fragments $c_k$, $z_{K-k}$ fragments with appropriate charge are summed.
Also, we entirely neglect the presence of HTR in the whole analysis, as it renders the whole procedure too complex: all fragments that could have been taken either for ETD or HTR products are considered to be purely ETD products. 
We then construct the \textit{pairing graph}, see Figure~\ref{fig::pairing graph}. 
The nodes correspond to different observed molecular species and store information on their total estimated intensity.
Special dummy nodes are added to denote the matching cofragments that had lost all their charge.
Our approach assumes that the only way ions can end up being undetected is solely through the total loss of charge.
Edges are drawn between $c$ and $z$ nodes with complimentary sequences if their total charge plus one (the fragmentation producing fragments takes away one charged) does not exceed that of the precursor chosen in MS1 stage of the experiment, $q_c + q_z + 1 \leq Q$. 
\begin{figure}[t]
    \includegraphics[width=.9\linewidth]{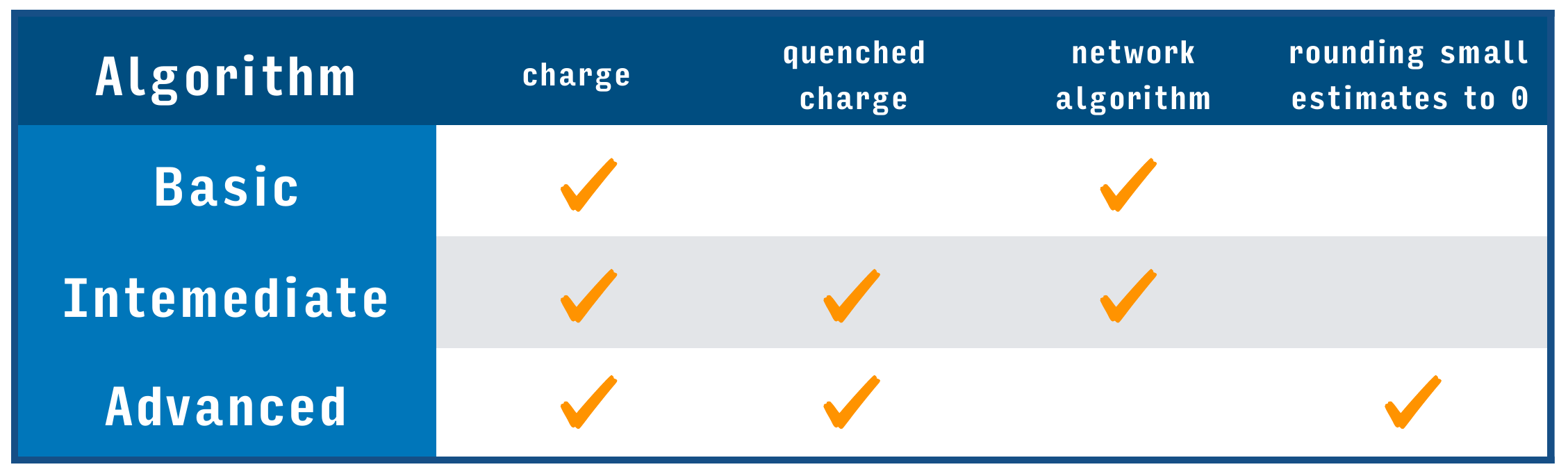}
    \caption{Summary of the proposed pairing algorithms.}\label{fig::algorithms}
\end{figure}

The pairing of fragments correspond to the redistribution of the estimated intensities $I$ in the nodes along the edges of the pairing graph. 
Assigning intensity to an edge diminishes the intensities in both end nodes by the same amount. 
All intensity must be assigned to some edges. 
Assigning intensity comes at a cost reflecting the number of reactions the pair of ions underwent during the whole experiment.
In the basic approach, fragments with charges $(q_c, q_z)$ together underwent $Q - 1 - q_c - q_z$ reactions.
The optimization task we are about to set up lets us forget the extra fragmentation count, fixing these costs at $Q - q_c - q_z$, equal to the total number of ETnoD and PTR reactions that both fragments underwent, $N_\text{ETnoD}^{cz} + N_\text{PTR}^{cz}$. 
Note that this equation holds also for pairings involving cofragments that entirely disappeared due to the loss of all charge. 

The pairing problem turns into an optimization problem where one wants to minimize the total number of reactions that could have produced the observed $c$ and $z$ fragments.
More specifically, we face a constrained linear optimization task: 
\begin{align}
	\min_{ I_{cz}:\,c \in \mathcal{A}_C,\,z \in \mathcal{A}_Z }
	\sum_{ \mbox{\tiny$\begin{matrix} 
		c \in \mathcal{A}_C \\
		z \in \mathcal{A}_Z
	\end{matrix}$} } (N^{cz}_\text{ETnoD} + N^{cz}_\text{PTR}) I_{cz}\label{eq::cost of pairing} \\
	\forall_{c\in\mathcal{O}_C}\,\, I_c = \sum_{z\in\mathcal{A_Z}} I_{cz},\,\,\forall_{z\in\mathcal{O}_Z}\,\, I_z = \sum_{c\in \mathcal{A_C}} I_{cz}. \label{eq::equality constraints}
\end{align}
Above, $\mathcal{O}_C$ and $\mathcal{O}_Z$ denote sets of observed $c$ and $z$ nodes, and $\mathcal{A}_C$ and $\mathcal{A}_Z$ additionally contain the unobserved cofragments. 

The above simplifies to a \textit{max flow} problem: subtract flows between observed fragments from both sides of equalities in \eqref{eq::equality constraints} and what results are the expressions for flows between observed fragments and their unobserved cofragments. 
Plugging these into Eq.~\eqref{eq::cost of pairing} and some simple algebra results in
\begin{align*}
	\max_{ I_{cz}:\,c \in \mathcal{A}_C,\,z \in \mathcal{A}_Z } 
	\sum_{\mbox{\tiny$\begin{matrix} 
		c \in \mathcal{O}_C \\
		z \in \mathcal{O}_Z
	\end{matrix}$}} I_{cz} \quad\text{s.t.}\\
	\forall_{c\in\mathcal{O}_C} I_c \geq \sum_{z\in\mathcal{O}_Z} I_{cz},\quad \forall_{z\in\mathcal{O}_Z} I_z \geq \sum_{c\in\mathcal{O}_C} I_{cz}.
\end{align*}
Of course, all flows $I_{cz}$ are non-negative.
To solve the max flow problem we use the Edmonds-Karp algorithm\cite{edmonds1972theoretical} as implemented in the {\tt NetworkX} Python module\cite{hagberg-2008-exploring}.

The solution to the above problem provides us with estimates of the total intensities of ions undergoing a specific type of fragmentation. 
In particular, this lets us estimate the probabilities of fragmentation along the protein. 
It also lets us estimate the probability with which the precursor will fragment. 
However, this setting does not offer any possibility to estimate the number of ETnoD and PTR reactions from fragments. 
These might become important in case of experiments where bigger and more charged substances are studied, or when much of the precursor ions reacted away, mostly through fragmentation.

To provide a solution to the above problems, we have developed another algorithm -- the \textit{intermediate} approach. 
In this approach we do not aggregate the intensities of observed ions with different quenched charges. 
As a result, the \textit{pairing graph} contains more nodes, both observed and dummy ones.
Have we followed the previous approach, then each observed fragment could match  several unobserved cofragments, all amounting to the same overall number of reactions but differing in specific numbers of ETnoD and PTR among them.
Unfortunately, the existence of many unobservable cofragments would prevent us from reducing the problem to a \textit{max flow} optimization, making it impossible to derive equations for all flows between observed and unobserved fragments. 
To solve this problem, we reduce the number of potential dummy nodes and combine them together.

The edges between existing fragments now convey information necessary to tell how many PTR and ETnoD reactions happened on both fragments throughout their history, including the period before any fragmentation ooccurred.
Similarly to equations \eqref{eq::q} and \eqref{eq::g}, the numbers of PTR and ETnoD reactions on a given pair of fragments characterized by charges $(q_c, q_z)$ and quenched charges $(g_c, g_z)$ follow equations
\begin{align}
	N_\text{PTR}	&= Q - 1 - q_c - q_z - g_c - g_z\notag\\
	N_\text{ETnoD}	&= q_c + q_z.\notag
\end{align}
Note that due to aggregation, the same cannot be said about edges between the observed and unobserved ions.
Otherwise said, if a mass spectrum does not contain \textit{pairable} fragments, then the only source of information on the numbers of ETnoD and PTR reactions can be obtained solely from the precursor products.

Finally, we investigated a third solution to the \textit{pairing problem}, the \textit{advanced} approach.
It includes the introduction of additional penalty terms to the cost function, 
\begin{equation*}
	\lambda_1 \sum_{
	 \mbox{\tiny$\begin{matrix} 
		c \in \mathcal{A}_C \\
		z \in \mathcal{A}_Z
	\end{matrix}$}} I_{cz} + 
	\lambda_2 \sum_{
	 \mbox{\tiny$\begin{matrix} 
		c \in \mathcal{A}_C \\
		z \in \mathcal{A}_Z
	\end{matrix}$}} I_{cz}^2.
\end{equation*}
Above, $\lambda_1$ corresponds to a lasso-type penalty and $\lambda_2$ - a ridge penalty. 
This approach was investigated mainly for its ability to automatically round the estimates of small flows to zero.
The above problem cannot be cast into the \textit{max flow} setting because of the quadratic terms in the cost function. 
For this reason, we use yet again the general purpose {\tt CVXOPT} solver.

\begin{algorithm}\scriptsize
	\caption{\textit{In silico} spectra generator}\label{alg::simulation}
	\begin{algorithmic}
		\State
		\INPUT
			\State A list $\mathcal{I}$ comprising $N$ precursor ions with a given charge $Q$ and sequence $F$.
			\State Probabilities of reactions $p_\text{PTR}, p_\text{ETnoD}, p_\text{ETD}$.
			\State Overall intensity $I$ of the process.
			\State Standard deviation of mass inaccuracy $\sigma$.
		\OUTPUT 
		\State A mass spectrum.
		\State
		\State Draw the placements of charges $q$ along the fasta sequence.
		\State Set experiment time to zero, $T = 0$.
		\While{ T < 1 }
			\State Increase T by a random time interval sampled from 
			\State \quad the exponential distribution with intensity $I\sum_i N_i q_i^2$.
			\State Extract ion $M$ from $\mathcal{I}$ with probability prop. to $N_i q_i^2$.  
			\State Draw R from PTR, ETnoD, and PTR,
			\State \quad with probabilities $p_\text{PTR}, p_\text{ETnoD}, p_\text{ETD}$.
			\If{ R = ETD }
				\If{ fragmentation occurred twice }
					\State Discard ion M.
				\Else
					\State Draw the fragmentation spot.
					\State Add fragments with $q>0$ to $\mathcal{I}$.
				\EndIf
			\Else	
				\State Reduce charge by one.
				\State Adjust the quenched charge.
				\State Add $M$ to list $\mathcal{I}$.
			\EndIf
		\EndWhile
		\ForAll{ $M$ in $\mathcal{I}$ }
			\State Randomly choose the isotopic variant of $M$.
			\State Blur its mass with gaussian noise.
		\EndFor
		\State bin the spectrum 
	\end{algorithmic}
\end{algorithm}

\section{Results and Discussion}

\noindent\textbf{In Silico results.} In order to test the entire workflow, we conducted \textit{in silico} experiments. 
A chemical process was simulated using a tailored Gillespie algorithm\cite{gillespie1977exact}, as described in Algorithm~\ref{alg::simulation}.
Briefly, the process generates a random series of three chemical reactions (PTR, ETnoD, and ETD; HTR is neglected) occurring in particular moments of time. The length of time intervals between reaction events is random and depends upon the number of charged ions at particular charge state, following \citet{McLuckey1999-su}. 


\begin{figure}[t]
	\centering
	\includegraphics[width=\linewidth]{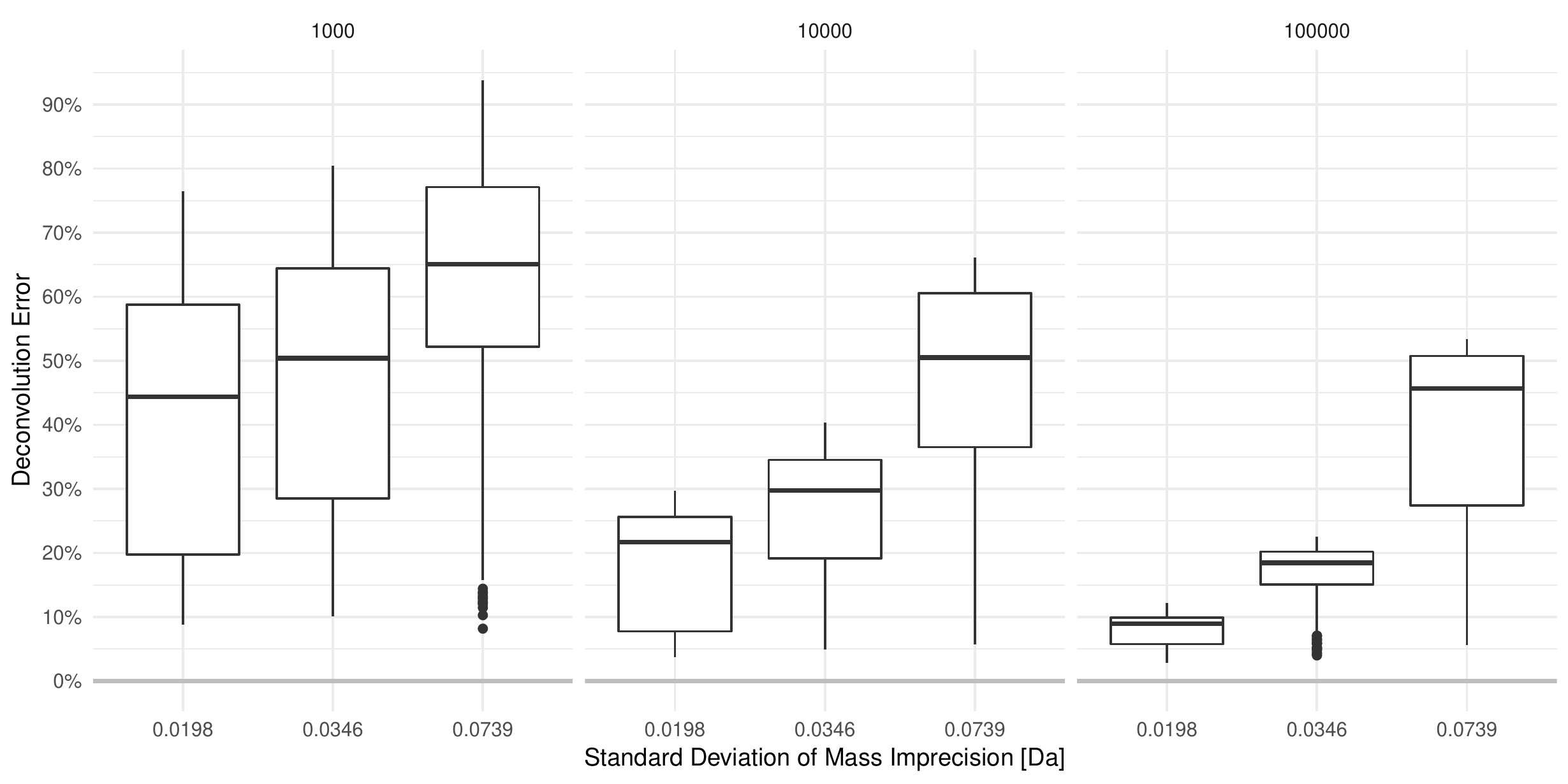}
	\caption{Error rates of the deconvolution procedure on \textit{in silico} data for different numbers of initial precursor ions (N = 1~000, 10~000, 100~000) and under different amounts of mass inaccuracy $sigma$ (on $x$ axis). The tolerance interval in \textsc{MassTodon} was set to $0.05$ [Da]. To measure error we sum the absolute differences of peak heights and normalize the result to the number of the precursor ions  (the result does not need to sum to 100\%).
	}
    \label{fig::in silico errors}
\end{figure}

{\tt MassTodon} was tested in various conditions: we checked all the combinations of settings of different initial numbers of precursors, $N = 1000, 10000$, or $100000$ ions, initial precursor charges $Q = 3, 6, 9,$ and $12$, three levels of the standard deviation of mass accuracy $\sigma$, and $12$ different sets of probabilities of reactions.

Strongest correlation with deconvolution error was noticed for spectra with low ion content and large mass inaccuracy, see Figure \ref{fig::in silico errors}.
The algorithm works best when there is enough ions to form a well sampled isotopic distribution (in case of our simulations -- 100~000 ions). 
In case of high-resolution mass data, when thousands of isotopologue peaks are present in the mass spectrum, it might be thus advisable to provide MassTodon with a spectrum binning results of several runs of the instrument. 
It is also vital not to underestimate the size of the tolerance interval. 
Of course, the above remarks are intrinsic to any peak assigning procedure that uses peak intensities, rather than relying solely on their mass over charge ratios. 

While running simulation descibed by Algorithm~\ref{alg::simulation} we store the numbers of each molecule $M$ drawn in the process. We have compared these numbers with the estimates of MassTodon to check the quality of the applied deconvolution procedures. 
Figure~\ref{fig::in silico errors} reports the obtained error rates. 

\begin{figure}[t]
	\centering
	\includegraphics[width=\linewidth]{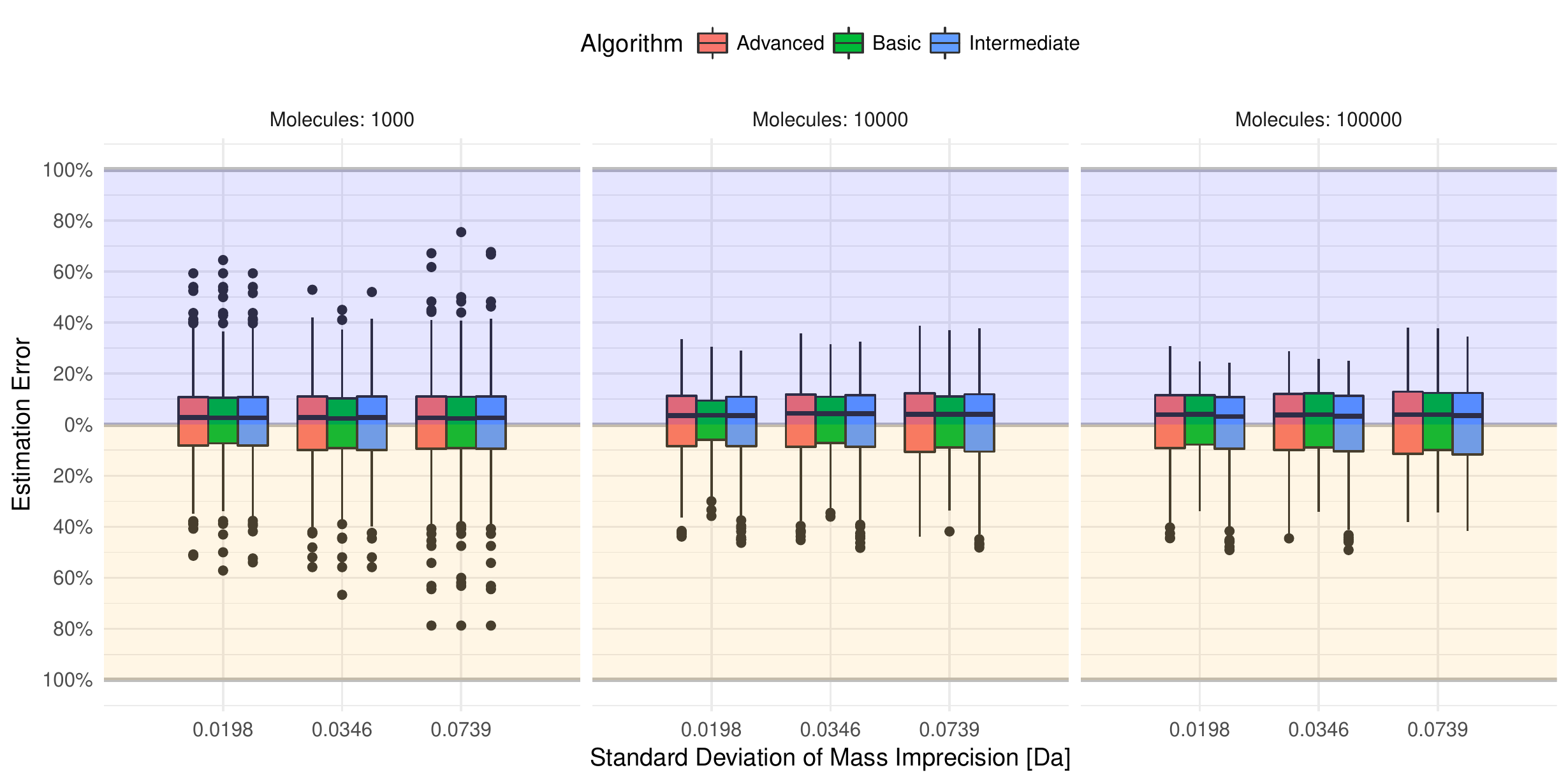}
	\caption{The distribution of distance between the estimates $(\hat{p}_\text{ETnoD}, \hat{p}_\text{PTR})$ and the true values $(p_\text{ETnoD}, p_\text{PTR})$ for different approaches we take, measured by the euclidean distance normalized to the maximal distance $\sqrt{2}$. Estimates in the blue regions favor PTR, while those in the yellow - ETnoD. The distributions are conditional on the number of initial precursor ions (N = 1~000, 10~000, 100~000) and different level of mass inaccuracy $sigma$ (on $x$ axis).}
    \label{fig::signed prob error estimate}
\end{figure}

Interestingly, the number of ions in the sample proves is of limited importance if one is interested in the estimation of the probabilities of ETnoD and PTR reactions, as shown in Figure~\ref{fig::signed prob error estimate}.
We note, that the parsimonious approach we have taken on average only slightly overestimates values of the true parameters, showing a preference towards the PTR reaction.
Note also, that the \textit{basic} approach to the pairing problem seems to offer estimates with the smallest variance. 

\noindent\textbf{Experimental results.} 
Mass spectra have been acquired for purified Substance P and ubiquitin as described in detail in the previous publications\cite{lermyte2015understanding,lermyte2015characterization}. 

\begin{figure*}[t]
    \begin{subfigure}[b]{\linewidth}
    	\begin{subfigure}[b]{0.33\linewidth}
	    	\centering
	        \caption{Mismatch and Fitting Errors}\label{fig::fitting errors on subP}\label{fig::fit errors}
	        \includegraphics[width=\textwidth]{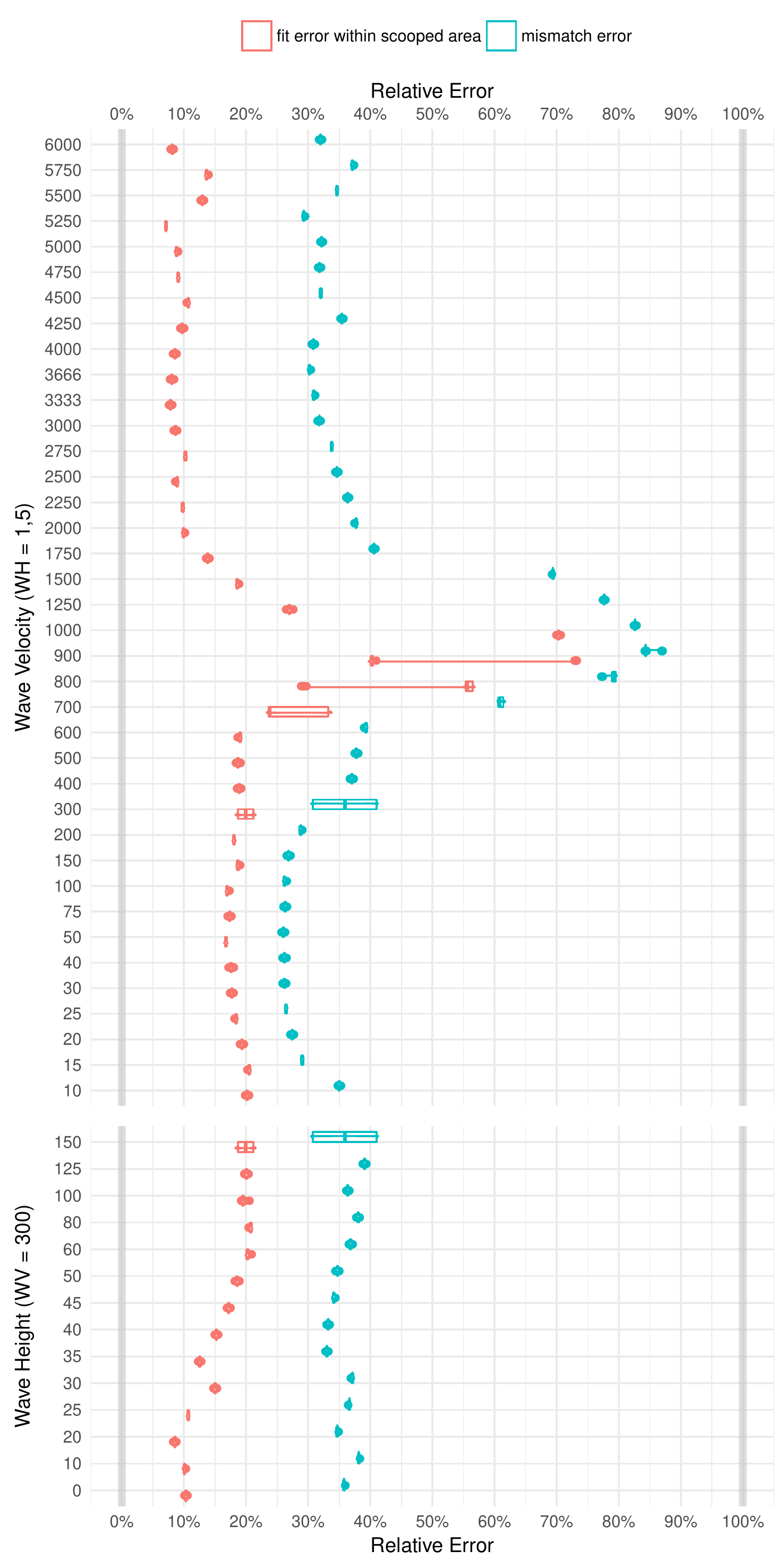}
        \end{subfigure}
        \begin{subfigure}[b]{0.66\linewidth}
	    	\centering
	    	\caption{Probabilities of Fragmentation}\label{fig::fragmentations}
	        \includegraphics[width=\textwidth]{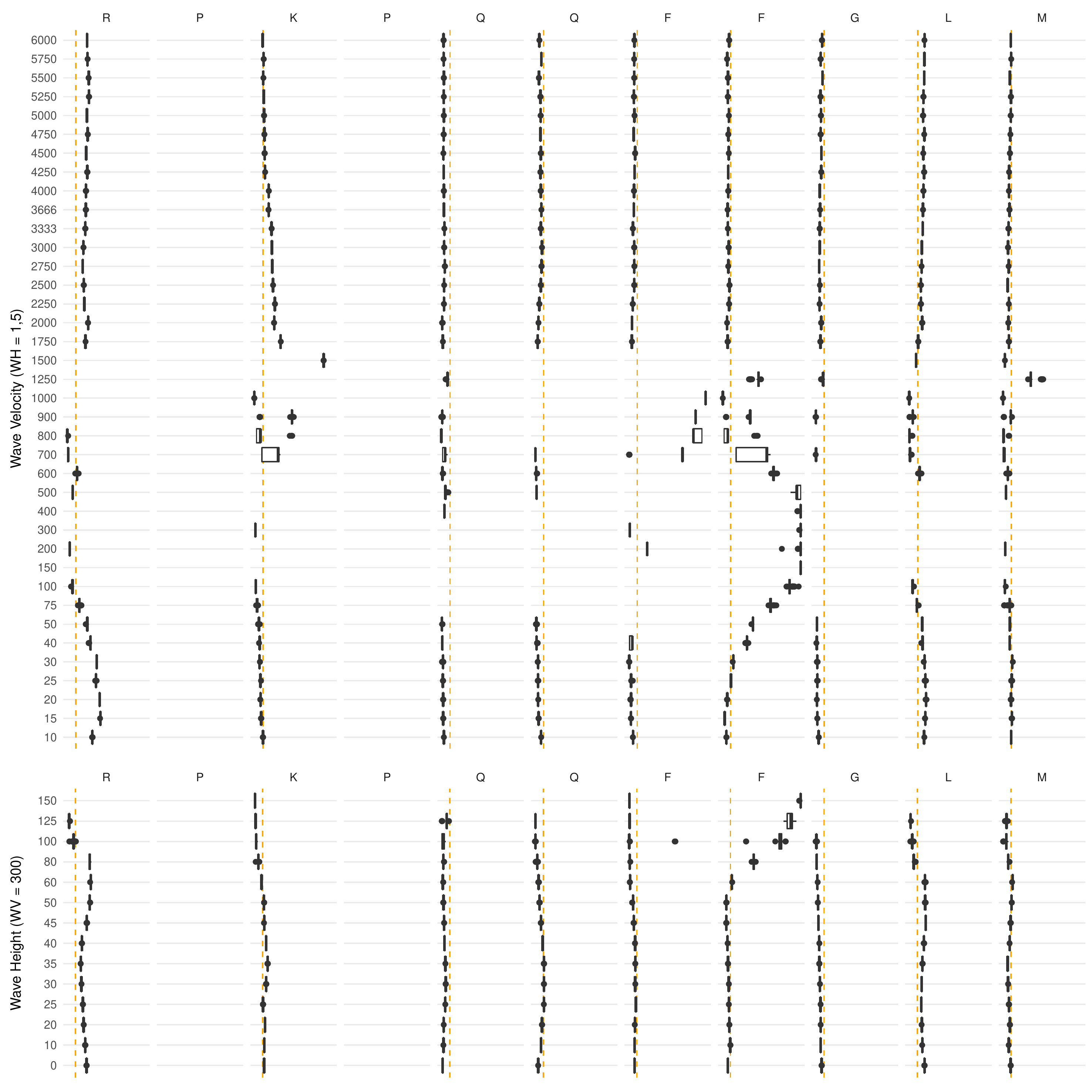}
        \end{subfigure}
    \end{subfigure}
    \begin{subfigure}[b]{\linewidth}
    	\includegraphics[width=\textwidth]{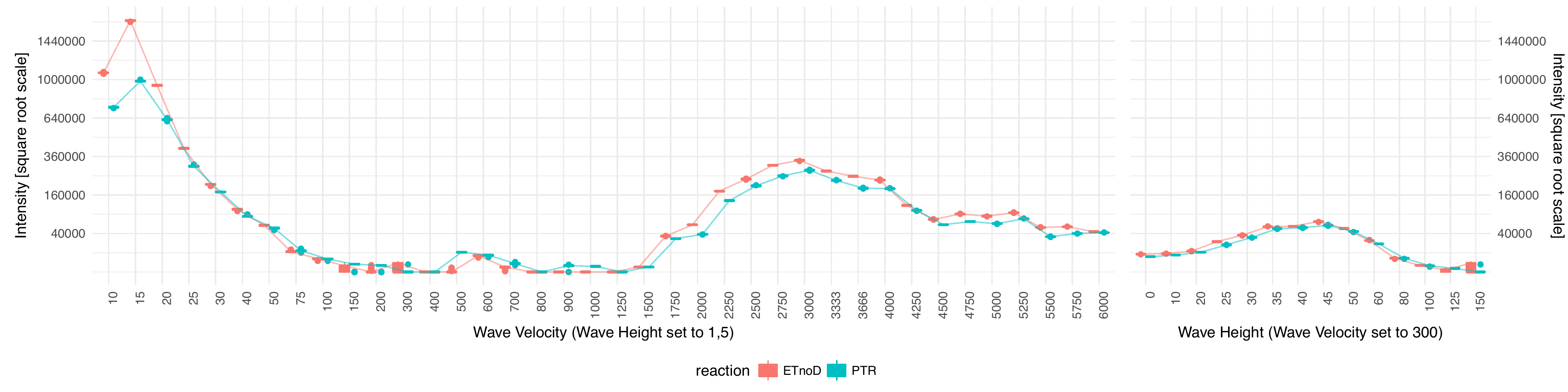}
    	\caption{Intensities of ETnoD and PTR}\label{fig::etnod ptr intensities}
    \end{subfigure}
    \caption{
    	Selected results of the \textsc{MassTodon} as run on substance P spectra. The instrumental settings were obtained for two two strips of settings in the two dimensional space comprising wave height and velocity. Results in (a) and (b) show bootstrap estimates (250 repetitions). Results in (c) contain additionally lines linking together the estimates obtained for the actual mass spectra. Figure (a) shows estimates of the mismatch error and the fitting error. Both are calculated using the normalized $l_1$ distance, $E(p,q) = \frac{ \sum_k |p_k - q_k| }{ \sum_k p_k + \sum_k q_k }$, where $p$ and $q$ are maps with keys $k$ (different m/z ranges) and values $p_k$ and $q_k$ (i.e. real intensities and their estimates). In case of the mismatch error, we compare in this way the estimated spectrum versus whole experimental mass spectrum, which includes peaks that are not among the studied reaction products. The fit error restricts this comparison to the regions of the mass spectrum that actually could be explained by some theoretical product of some reaction. Figure (b) shows estimates of the probabilities of fragmentation along the backbone of substance P, whose amino sequence is RPKPQQFFGLM. Fragmentation on prolines (P) is deemed highly unlikely due to the ring structure of this amino acid. The vertical orange dashed lines correspond to probability equal to $1/9$, which would be attained assuming a fully uniform probability of fragmentation. Figure (c) shows the estimates of the intensity of the ETnoD and PTR reactions. Values of intensities in the $y$ axis have been transformed by a square root scaling in order to expose the behaviour of the lower estimates.
    }\label{fig::substance P wall}
\end{figure*}

\begin{figure*}[t]    
	\centering
    \includegraphics[width=\textwidth]{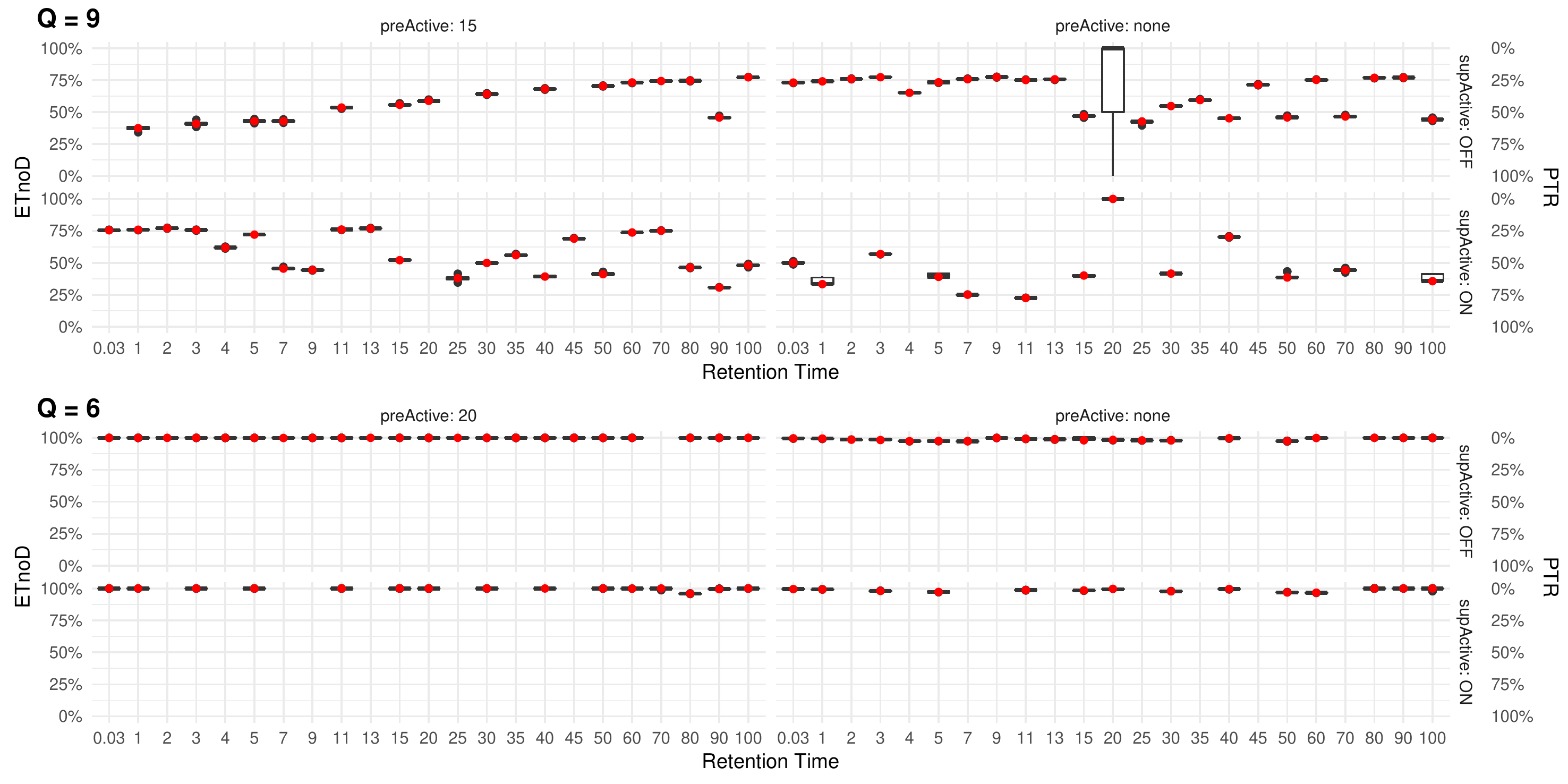}
    \caption{Estimates of the probabilities of ETnoD and PTR conditional on one of these events happening. Red dots correspond to estimates performed on real data. The black box plots, mostly extremely narrow, correspond to 250 sample bootstrap estimates. Precursor charge $Q$ is shown in top-left parts of the panels. Each panel is subdivided into subpanels corresponding to different experimental settings. \textit{Attention:} left panels correspond to different levels of preactivations. For $Q=9$ the energy of preactivation was set to 15, while for $Q = 6$ to 20. The $x$ axis shows the retention time RT, while the $y$ axis shows the percentual content of the ETnoD and PTR reactions. 
    For the spectrum gathered at $Q=9$ and RT = 20, without pre-activation and without the supplementary activation, there were no ions found that could undergo ETnoD or PTR in the real spectrum under the given threshold on the intensity (results contain the $95\%$ of the highest peak in that spectrum), so the red dot is missing. 
    During the bootstrap procedure, small percentages of peaks apparently corresponding to PTR or ETnoD products appeared above that threshold, leading to the 
    }\label{fig::ubi wall}
\end{figure*}

The outcomes of \textsc{MassTodon} can be used to compare more easily mass spectra gathered under different instrumental settings. 
Figures~\ref{fig::substance P wall} and \ref{fig::ubi wall} explore the differences and similarities of the information conveyed in different mass spectra, including their percentual content of products of all studied reactions, the probabilities of fragmentation, and intensities and probabilities of the ETnoD and PTR reactions.

\begin{figure}[t]
	\centering
	\includegraphics[width=\linewidth]{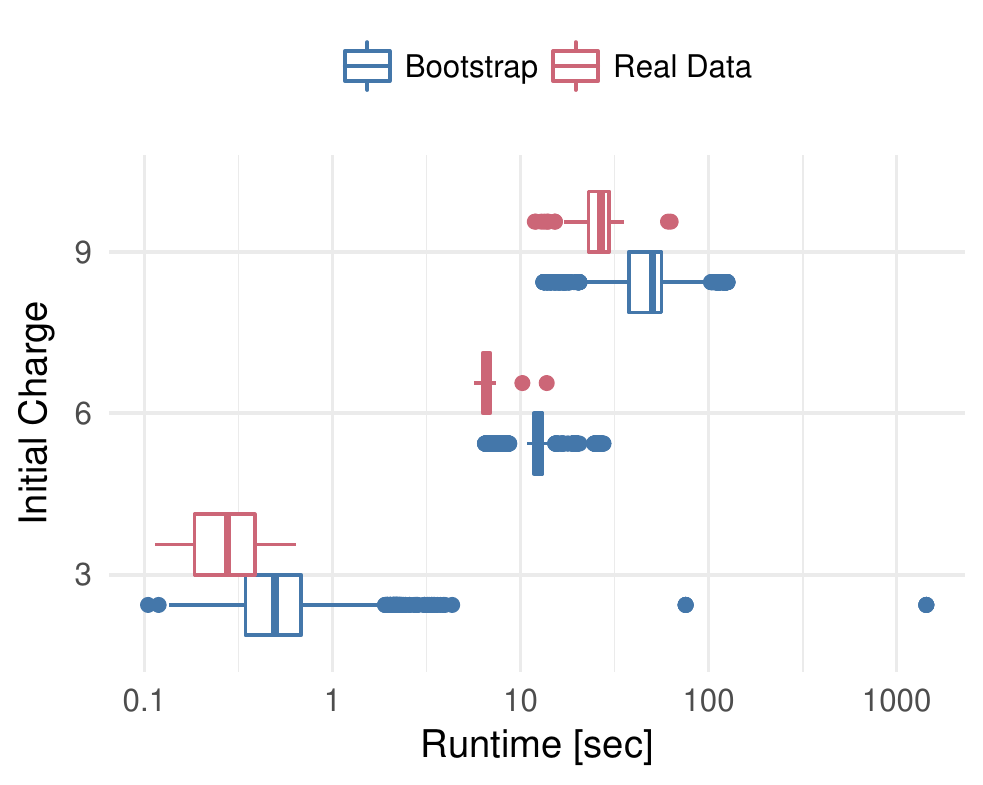}
	\caption{{\tt MassTodonPy} runtime distribution. The analysis contains all the stages of the algorithm, including running all three \textit{pairing algorithms}. The 3+ precursors correspond to substance P spectra; other results are obtained for ubiquitin. Usually, it takes more time to process a spectrum randomly reshuffled by bootstrap than the original version. Runtimes were obtained using the sequential version of the algorithm, which solves the \textit{deconvolution problems} one after another. It is possible to reduce this time for larger problems using the multiprocessing option.}
    \label{fig::runtime}
\end{figure}

\textsc{MassTodon} provides point estimates of the above parameters. 
Given that the analysis of one spectrum is reasonably fast (check Figure~\ref{fig::runtime}) we decided to rely on bootstrap procedures\cite{efron1994introduction,wasserman2013all} to estimate the standard deviations of the above parameters.
In particular, each mass spectrum was randomly reshuffled multiple times. 
We assume that each bootstrap spectrum to be composed out of $N$ ions. 
The m/z ratios of these ions were then independently drawn among the original ratios, with  probabilities equal to the heights of the corresponding peaks, normalized to the total ion current. 
The number of observed molecules in the spectrum $N$ is not truly known in advance. 
In our simulations, we assumed that the whole spectrum consist of around 100~000 molecules.  
We draw 250 random spectra for each real one and run \textsc{MassTodon} on each one of them.

Figure \ref{fig::fitting errors on subP} shows the overall fitting quality in case of the substance P spectra. On average, the products of the considered reactions cannot explain on average between 30\% to 40\% of the mass spectrum.
Shifting our attention only to those regions of the mass spectrum fall within the range of any potential product, the error estimates drops in a range between 10 to 20\%.   
Note that for spectra gather at wave height fixed at 150 and wave velocity between 700 to 1500 the errors grow significantly. 

Figure~\ref{fig::fragmentations} presents the estimates of probabilities of fragmentation for substance P. Interestingly, the probabilities are almost constant across different experimental settings.
They are also almost uniformly distributed along the possible fragmentation sites (proline not being one of them).
This is what would be expected of a small molecule, like substance P, with a trivial tertiary structure.
Again, significant departures from this pattern emerge in the same region of wave velocity. 

Figure~\ref{fig::etnod ptr intensities} seems to shed some light on the nature of these anomalies. 
It presents the estimates of the intensity of ions that underwent ETnoD and PTR, which is a proxy for the number of these events to happen on the molecules of substance P within the sample. 
In particular, it can be noted that the range of wave velocity between 700 to 1500 contains a particularly small amount of ions that could have been prescribed to ETnoD or PTR. 
By comparison, all estimates where these intensities were above 40~000 show a much smaller amount of variance.  
Note also, that Figure~\ref{fig::etnod ptr intensities} suggests that the relative ratios of ETnoD and PTR remain stable under most experimental settings, with the exception of small wave velocities. 
These ratios can be interpreted as relative probabilities of the ETnoD and PTR reactions, conditional on one of the reaction happening.

Interestingly, a similar pattern reemerges in case of mass spectra of ubiquitin, as shown in Figure~\ref{fig::ubi wall}.
In case of the mass spectra where the filtered precursor molecule was bearing 6 charges, the ETnoD vastly dominates over PTR. 
In one of our previous papers\cite{lermyte2015understanding} we show, that this might be related to the insufficient capability of only 6 protons to induce enough denaturation of the protein within the instrument.
In other words, the fragmentation cannot happen because the two fragments wrap around each other, giving rise to a higher percentage of the ETnoD products.

\section{Conclusion}
As high-performance mass spectrometers and the use of ExD methods become more prevalent, there will be an increasing demand for software methods to assist in processing the resulting, considerable amounts of data. 
Here, we have presented a user-friendly software package to analyze high-resolution ETD data, deconvolute isotope distributions, and infer information about various competing reaction pathways occurring under ETD conditions. 

Future work will focus on casting the entire framework into a Bayesian setting, in order to provide the user with better understanding of the uncertainties of the estimates and potential correlations of results. 
In particular, the user might be interested to what degree some parts of the mass spectrum could be alternatively explained by other substances. 
Obtaining such information could be done by looking at the joint distribution of the counts of molecules that compete for the explanation of a given part of the spectrum.

Moreover, it would be interesting to free the user from the need to specify the tolerance parameter. 
This should be obtained automatically and potentially vary in different ranges of the m/z half line.

The implementation of the {\tt MassTodon} algorithm is freely available for downloads from the Python Package Index.
Installation instructions and documentation can be found at \href{http://masstodonpy.readthedocs.io}{\tt readthedocs}.
Source code is available for download from \href{https://matteolacki.github.io/MassTodonPy/}{\tt github}.
The software is distributed under the terms of the GNU GPL V3 public license.

\noindent\textbf{Acknowledgements.}
We would like to thank Michał Aleksander Ciach for his help in implementing the \textit{in silico} simulator.
Finally, we would like to thank dr Piotr Dittwald for his continuous support.

This work is supported by Polish NCN grants 2014/12/W/ST5/00592, 2015/17/N/ST6/03565 and partially by the Flemish SBO grant InSPECtor, 120025, IWT.

{\scriptsize\bibliography{bib/frederik,bib/spectrometry,bib/statistics,bib/algorithmics}}

\providecommand{\latin}[1]{#1}
\providecommand*\mcitethebibliography{\thebibliography}
\csname @ifundefined\endcsname{endmcitethebibliography}
  {\let\endmcitethebibliography\endthebibliography}{}
\begin{mcitethebibliography}{51}
\providecommand*\natexlab[1]{#1}
\providecommand*\mciteSetBstSublistMode[1]{}
\providecommand*\mciteSetBstMaxWidthForm[2]{}
\providecommand*\mciteBstWouldAddEndPuncttrue
  {\def\EndOfBibitem{\unskip.}}
\providecommand*\mciteBstWouldAddEndPunctfalse
  {\let\EndOfBibitem\relax}
\providecommand*\mciteSetBstMidEndSepPunct[3]{}
\providecommand*\mciteSetBstSublistLabelBeginEnd[3]{}
\providecommand*\EndOfBibitem{}
\mciteSetBstSublistMode{f}
\mciteSetBstMaxWidthForm{subitem}{(\alph{mcitesubitemcount})}
\mciteSetBstSublistLabelBeginEnd
  {\mcitemaxwidthsubitemform\space}
  {\relax}
  {\relax}

\bibitem[Zubarev \latin{et~al.}(1998)Zubarev, Kelleher, and
  McLafferty]{zubarev1998electron}
Zubarev,~R.~A.; Kelleher,~N.~L.; McLafferty,~F.~W. \emph{Journal of the
  American Chemical Society} \textbf{1998}, \emph{120}, 3265--3266\relax
\mciteBstWouldAddEndPuncttrue
\mciteSetBstMidEndSepPunct{\mcitedefaultmidpunct}
{\mcitedefaultendpunct}{\mcitedefaultseppunct}\relax
\EndOfBibitem
\bibitem[Syka \latin{et~al.}(2004)Syka, Coon, Schroeder, Shabanowitz, and
  Hunt]{Syka2004PeptideAndProtein}
Syka,~J. E.~P.; Coon,~J.~J.; Schroeder,~M.~J.; Shabanowitz,~J.; Hunt,~D.~F.
  \emph{Proceedings of the National Academy of Sciences of the United States of
  America} \textbf{2004}, \emph{101}, 9528--9533\relax
\mciteBstWouldAddEndPuncttrue
\mciteSetBstMidEndSepPunct{\mcitedefaultmidpunct}
{\mcitedefaultendpunct}{\mcitedefaultseppunct}\relax
\EndOfBibitem
\bibitem[Garcia \latin{et~al.}(2007)Garcia, Shabanowitz, and
  Hunt]{garcia2007characterization}
Garcia,~B.~A.; Shabanowitz,~J.; Hunt,~D.~F. \emph{Current opinion in chemical
  biology} \textbf{2007}, \emph{11}, 66--73\relax
\mciteBstWouldAddEndPuncttrue
\mciteSetBstMidEndSepPunct{\mcitedefaultmidpunct}
{\mcitedefaultendpunct}{\mcitedefaultseppunct}\relax
\EndOfBibitem
\bibitem[H{\aa}kansson \latin{et~al.}(2001)H{\aa}kansson, Cooper, Emmett,
  Costello, Marshall, and Nilsson]{haakansson2001electron}
H{\aa}kansson,~K.; Cooper,~H.~J.; Emmett,~M.~R.; Costello,~C.~E.;
  Marshall,~A.~G.; Nilsson,~C.~L. \emph{Analytical chemistry} \textbf{2001},
  \emph{73}, 4530--4536\relax
\mciteBstWouldAddEndPuncttrue
\mciteSetBstMidEndSepPunct{\mcitedefaultmidpunct}
{\mcitedefaultendpunct}{\mcitedefaultseppunct}\relax
\EndOfBibitem
\bibitem[Ayaz-Guner \latin{et~al.}(2009)Ayaz-Guner, Zhang, Li, Walker, and
  Ge]{ayaz2009vivo}
Ayaz-Guner,~S.; Zhang,~J.; Li,~L.; Walker,~J.~W.; Ge,~Y. \emph{Biochemistry}
  \textbf{2009}, \emph{48}, 8161--8170\relax
\mciteBstWouldAddEndPuncttrue
\mciteSetBstMidEndSepPunct{\mcitedefaultmidpunct}
{\mcitedefaultendpunct}{\mcitedefaultseppunct}\relax
\EndOfBibitem
\bibitem[Ge \latin{et~al.}(2009)Ge, Rybakova, Xu, and Moss]{ge2009top}
Ge,~Y.; Rybakova,~I.~N.; Xu,~Q.; Moss,~R.~L. \emph{Proceedings of the National
  Academy of Sciences} \textbf{2009}, \emph{106}, 12658--12663\relax
\mciteBstWouldAddEndPuncttrue
\mciteSetBstMidEndSepPunct{\mcitedefaultmidpunct}
{\mcitedefaultendpunct}{\mcitedefaultseppunct}\relax
\EndOfBibitem
\bibitem[Tsybin \latin{et~al.}(2011)Tsybin, Fornelli, Stoermer, Luebeck, Parra,
  Nallet, Wurm, and Hartmer]{tsybin2011structural}
Tsybin,~Y.~O.; Fornelli,~L.; Stoermer,~C.; Luebeck,~M.; Parra,~J.; Nallet,~S.;
  Wurm,~F.~M.; Hartmer,~R. \emph{Analytical chemistry} \textbf{2011},
  \emph{83}, 8919--8927\relax
\mciteBstWouldAddEndPuncttrue
\mciteSetBstMidEndSepPunct{\mcitedefaultmidpunct}
{\mcitedefaultendpunct}{\mcitedefaultseppunct}\relax
\EndOfBibitem
\bibitem[Fornelli \latin{et~al.}(2012)Fornelli, Damoc, Thomas, Kelleher,
  Aizikov, Denisov, Makarov, and Tsybin]{fornelli2012analysis}
Fornelli,~L.; Damoc,~E.; Thomas,~P.~M.; Kelleher,~N.~L.; Aizikov,~K.;
  Denisov,~E.; Makarov,~A.; Tsybin,~Y.~O. \emph{Molecular \& Cellular
  Proteomics} \textbf{2012}, \emph{11}, 1758--1767\relax
\mciteBstWouldAddEndPuncttrue
\mciteSetBstMidEndSepPunct{\mcitedefaultmidpunct}
{\mcitedefaultendpunct}{\mcitedefaultseppunct}\relax
\EndOfBibitem
\bibitem[Cournoyer \latin{et~al.}(2005)Cournoyer, Pittman, Ivleva, Fallows,
  Waskell, Costello, and O'Connor]{cournoyer2005deamidation}
Cournoyer,~J.~J.; Pittman,~J.~L.; Ivleva,~V.~B.; Fallows,~E.; Waskell,~L.;
  Costello,~C.~E.; O'Connor,~P.~B. \emph{Protein science} \textbf{2005},
  \emph{14}, 452--463\relax
\mciteBstWouldAddEndPuncttrue
\mciteSetBstMidEndSepPunct{\mcitedefaultmidpunct}
{\mcitedefaultendpunct}{\mcitedefaultseppunct}\relax
\EndOfBibitem
\bibitem[Li \latin{et~al.}(2010)Li, Lin, and O’Connor]{li2010glutamine}
Li,~X.; Lin,~C.; O’Connor,~P.~B. \emph{Analytical chemistry} \textbf{2010},
  \emph{82}, 3606--3615\relax
\mciteBstWouldAddEndPuncttrue
\mciteSetBstMidEndSepPunct{\mcitedefaultmidpunct}
{\mcitedefaultendpunct}{\mcitedefaultseppunct}\relax
\EndOfBibitem
\bibitem[Xie \latin{et~al.}(2006)Xie, Zhang, Yin, and Loo]{xie2006top}
Xie,~Y.; Zhang,~J.; Yin,~S.; Loo,~J.~A. \emph{Journal of the American Chemical
  Society} \textbf{2006}, \emph{128}, 14432--14433\relax
\mciteBstWouldAddEndPuncttrue
\mciteSetBstMidEndSepPunct{\mcitedefaultmidpunct}
{\mcitedefaultendpunct}{\mcitedefaultseppunct}\relax
\EndOfBibitem
\bibitem[Jackson \latin{et~al.}(2009)Jackson, Dutta, and Woods]{jackson2009use}
Jackson,~S.~N.; Dutta,~S.; Woods,~A.~S. \emph{Journal of the American Society
  for Mass Spectrometry} \textbf{2009}, \emph{20}, 176--179\relax
\mciteBstWouldAddEndPuncttrue
\mciteSetBstMidEndSepPunct{\mcitedefaultmidpunct}
{\mcitedefaultendpunct}{\mcitedefaultseppunct}\relax
\EndOfBibitem
\bibitem[Yin and Loo(2010)Yin, and Loo]{yin2010elucidating}
Yin,~S.; Loo,~J.~A. \emph{Journal of the American Society for Mass
  Spectrometry} \textbf{2010}, \emph{21}, 899--907\relax
\mciteBstWouldAddEndPuncttrue
\mciteSetBstMidEndSepPunct{\mcitedefaultmidpunct}
{\mcitedefaultendpunct}{\mcitedefaultseppunct}\relax
\EndOfBibitem
\bibitem[G{\"o}th \latin{et~al.}(2016)G{\"o}th, Lermyte, Schmitt, Warnke, von
  Helden, Sobott, and Pagel]{goth2016gas}
G{\"o}th,~M.; Lermyte,~F.; Schmitt,~X.~J.; Warnke,~S.; von Helden,~G.;
  Sobott,~F.; Pagel,~K. \emph{Analyst} \textbf{2016}, \emph{141},
  5502--5510\relax
\mciteBstWouldAddEndPuncttrue
\mciteSetBstMidEndSepPunct{\mcitedefaultmidpunct}
{\mcitedefaultendpunct}{\mcitedefaultseppunct}\relax
\EndOfBibitem
\bibitem[Breuker \latin{et~al.}(2002)Breuker, Oh, Horn, Cerda, and
  McLafferty]{breuker2002detailed}
Breuker,~K.; Oh,~H.; Horn,~D.~M.; Cerda,~B.~A.; McLafferty,~F.~W. \emph{Journal
  of the American Chemical Society} \textbf{2002}, \emph{124}, 6407--6420\relax
\mciteBstWouldAddEndPuncttrue
\mciteSetBstMidEndSepPunct{\mcitedefaultmidpunct}
{\mcitedefaultendpunct}{\mcitedefaultseppunct}\relax
\EndOfBibitem
\bibitem[Oh \latin{et~al.}(2002)Oh, Breuker, Sze, Ge, Carpenter, and
  McLafferty]{oh2002secondary}
Oh,~H.; Breuker,~K.; Sze,~S.~K.; Ge,~Y.; Carpenter,~B.~K.; McLafferty,~F.~W.
  \emph{Proceedings of the National Academy of Sciences} \textbf{2002},
  \emph{99}, 15863--15868\relax
\mciteBstWouldAddEndPuncttrue
\mciteSetBstMidEndSepPunct{\mcitedefaultmidpunct}
{\mcitedefaultendpunct}{\mcitedefaultseppunct}\relax
\EndOfBibitem
\bibitem[Skinner \latin{et~al.}(2012)Skinner, McLafferty, and
  Breuker]{skinner2012ubiquitin}
Skinner,~O.~S.; McLafferty,~F.~W.; Breuker,~K. \emph{Journal of the American
  Society for Mass Spectrometry} \textbf{2012}, \emph{23}, 1011--1014\relax
\mciteBstWouldAddEndPuncttrue
\mciteSetBstMidEndSepPunct{\mcitedefaultmidpunct}
{\mcitedefaultendpunct}{\mcitedefaultseppunct}\relax
\EndOfBibitem
\bibitem[Skinner \latin{et~al.}(2013)Skinner, Breuker, and
  McLafferty]{skinner2013charge}
Skinner,~O.~S.; Breuker,~K.; McLafferty,~F.~W. \emph{Journal of the American
  Society for Mass Spectrometry} \textbf{2013}, \emph{24}, 807--810\relax
\mciteBstWouldAddEndPuncttrue
\mciteSetBstMidEndSepPunct{\mcitedefaultmidpunct}
{\mcitedefaultendpunct}{\mcitedefaultseppunct}\relax
\EndOfBibitem
\bibitem[Zhang \latin{et~al.}(2011)Zhang, Cui, Wen, Blankenship, and
  Gross]{zhang2011native}
Zhang,~H.; Cui,~W.; Wen,~J.; Blankenship,~R.~E.; Gross,~M.~L. \emph{Analytical
  chemistry} \textbf{2011}, \emph{83}, 5598--5606\relax
\mciteBstWouldAddEndPuncttrue
\mciteSetBstMidEndSepPunct{\mcitedefaultmidpunct}
{\mcitedefaultendpunct}{\mcitedefaultseppunct}\relax
\EndOfBibitem
\bibitem[Zhang \latin{et~al.}(2013)Zhang, Cui, and Gross]{zhang2013native}
Zhang,~H.; Cui,~W.; Gross,~M.~L. \emph{International journal of mass
  spectrometry} \textbf{2013}, \emph{354}, 288--291\relax
\mciteBstWouldAddEndPuncttrue
\mciteSetBstMidEndSepPunct{\mcitedefaultmidpunct}
{\mcitedefaultendpunct}{\mcitedefaultseppunct}\relax
\EndOfBibitem
\bibitem[Zhang \latin{et~al.}(2014)Zhang, Browne, and
  Vachet]{zhang2014exploring}
Zhang,~Z.; Browne,~S.~J.; Vachet,~R.~W. \emph{Journal of the American Society
  for Mass Spectrometry} \textbf{2014}, \emph{25}, 604--613\relax
\mciteBstWouldAddEndPuncttrue
\mciteSetBstMidEndSepPunct{\mcitedefaultmidpunct}
{\mcitedefaultendpunct}{\mcitedefaultseppunct}\relax
\EndOfBibitem
\bibitem[Lermyte \latin{et~al.}(2014)Lermyte, Konijnenberg, Williams, Brown,
  Valkenborg, and Sobott]{lermyte2014etd}
Lermyte,~F.; Konijnenberg,~A.; Williams,~J.~P.; Brown,~J.~M.; Valkenborg,~D.;
  Sobott,~F. \emph{Journal of The American Society for Mass Spectrometry}
  \textbf{2014}, \emph{25}, 343--350\relax
\mciteBstWouldAddEndPuncttrue
\mciteSetBstMidEndSepPunct{\mcitedefaultmidpunct}
{\mcitedefaultendpunct}{\mcitedefaultseppunct}\relax
\EndOfBibitem
\bibitem[Lermyte and Sobott(2015)Lermyte, and Sobott]{lermyte2015electron}
Lermyte,~F.; Sobott,~F. \emph{Proteomics} \textbf{2015}, \emph{15},
  2813--2822\relax
\mciteBstWouldAddEndPuncttrue
\mciteSetBstMidEndSepPunct{\mcitedefaultmidpunct}
{\mcitedefaultendpunct}{\mcitedefaultseppunct}\relax
\EndOfBibitem
\bibitem[Zhang \latin{et~al.}(2016)Zhang, Cui, Wecksler, Zhang, Molina,
  Deperalta, and Gross]{zhang2016native}
Zhang,~Y.; Cui,~W.; Wecksler,~A.~T.; Zhang,~H.; Molina,~P.; Deperalta,~G.;
  Gross,~M.~L. \emph{Journal of The American Society for Mass Spectrometry}
  \textbf{2016}, \emph{27}, 1139--1142\relax
\mciteBstWouldAddEndPuncttrue
\mciteSetBstMidEndSepPunct{\mcitedefaultmidpunct}
{\mcitedefaultendpunct}{\mcitedefaultseppunct}\relax
\EndOfBibitem
\bibitem[Lermyte \latin{et~al.}(2017)Lermyte, {\L}{\k{a}}cki, Valkenborg,
  Gambin, and Sobott]{lermyte2017conformational}
Lermyte,~F.; {\L}{\k{a}}cki,~M.~K.; Valkenborg,~D.; Gambin,~A.; Sobott,~F.
  \emph{Journal of The American Society for Mass Spectrometry} \textbf{2017},
  \emph{28}, 69--76\relax
\mciteBstWouldAddEndPuncttrue
\mciteSetBstMidEndSepPunct{\mcitedefaultmidpunct}
{\mcitedefaultendpunct}{\mcitedefaultseppunct}\relax
\EndOfBibitem
\bibitem[Ture\v{c}ek(2003)]{turecek2003n}
Ture\v{c}ek,~F. \emph{Journal of the American Chemical Society} \textbf{2003},
  \emph{125}, 5954--5963\relax
\mciteBstWouldAddEndPuncttrue
\mciteSetBstMidEndSepPunct{\mcitedefaultmidpunct}
{\mcitedefaultendpunct}{\mcitedefaultseppunct}\relax
\EndOfBibitem
\bibitem[Ture\v{c}ek and Syrstad(2003)Ture\v{c}ek, and
  Syrstad]{turecek2003mechanism}
Ture\v{c}ek,~F.; Syrstad,~E.~A. \emph{Journal of the American Chemical Society}
  \textbf{2003}, \emph{125}, 3353--3369\relax
\mciteBstWouldAddEndPuncttrue
\mciteSetBstMidEndSepPunct{\mcitedefaultmidpunct}
{\mcitedefaultendpunct}{\mcitedefaultseppunct}\relax
\EndOfBibitem
\bibitem[Chung and Ture{\v{c}}ek(2010)Chung, and
  Ture{\v{c}}ek]{chung2010backbone}
Chung,~T.~W.; Ture{\v{c}}ek,~F. \emph{Journal of the American Society for Mass
  Spectrometry} \textbf{2010}, \emph{21}, 1279--1295\relax
\mciteBstWouldAddEndPuncttrue
\mciteSetBstMidEndSepPunct{\mcitedefaultmidpunct}
{\mcitedefaultendpunct}{\mcitedefaultseppunct}\relax
\EndOfBibitem
\bibitem[Horn \latin{et~al.}(2000)Horn, Zubarev, and
  McLafferty]{horn2000automated}
Horn,~D.~M.; Zubarev,~R.~A.; McLafferty,~F.~W. \emph{Journal of the American
  Society for Mass Spectrometry} \textbf{2000}, \emph{11}, 320--332\relax
\mciteBstWouldAddEndPuncttrue
\mciteSetBstMidEndSepPunct{\mcitedefaultmidpunct}
{\mcitedefaultendpunct}{\mcitedefaultseppunct}\relax
\EndOfBibitem
\bibitem[Guner \latin{et~al.}(2014)Guner, Close, Cai, Zhang, Peng, Gregorich,
  and Ge]{guner2014mash}
Guner,~H.; Close,~P.~L.; Cai,~W.; Zhang,~H.; Peng,~Y.; Gregorich,~Z.~R.; Ge,~Y.
  \emph{Journal of The American Society for Mass Spectrometry} \textbf{2014},
  \emph{25}, 464--470\relax
\mciteBstWouldAddEndPuncttrue
\mciteSetBstMidEndSepPunct{\mcitedefaultmidpunct}
{\mcitedefaultendpunct}{\mcitedefaultseppunct}\relax
\EndOfBibitem
\bibitem[Cai \latin{et~al.}(2016)Cai, Guner, Gregorich, Chen, Ayaz-Guner, Peng,
  Valeja, Liu, and Ge]{cai2016mash}
Cai,~W.; Guner,~H.; Gregorich,~Z.~R.; Chen,~A.~J.; Ayaz-Guner,~S.; Peng,~Y.;
  Valeja,~S.~G.; Liu,~X.; Ge,~Y. \emph{Molecular \& Cellular Proteomics}
  \textbf{2016}, \emph{15}, 703--714\relax
\mciteBstWouldAddEndPuncttrue
\mciteSetBstMidEndSepPunct{\mcitedefaultmidpunct}
{\mcitedefaultendpunct}{\mcitedefaultseppunct}\relax
\EndOfBibitem
\bibitem[Mayampurath \latin{et~al.}(2008)Mayampurath, Jaitly, Purvine, Monroe,
  Auberry, Adkins, and Smith]{mayampurath2008deconmsn}
Mayampurath,~A.~M.; Jaitly,~N.; Purvine,~S.~O.; Monroe,~M.~E.; Auberry,~K.~J.;
  Adkins,~J.~N.; Smith,~R.~D. \emph{Bioinformatics} \textbf{2008}, \emph{24},
  1021--1023\relax
\mciteBstWouldAddEndPuncttrue
\mciteSetBstMidEndSepPunct{\mcitedefaultmidpunct}
{\mcitedefaultendpunct}{\mcitedefaultseppunct}\relax
\EndOfBibitem
\bibitem[Jaitly \latin{et~al.}(2009)Jaitly, Mayampurath, Littlefield, Adkins,
  Anderson, and Smith]{jaitly2009decon2ls}
Jaitly,~N.; Mayampurath,~A.; Littlefield,~K.; Adkins,~J.~N.; Anderson,~G.~A.;
  Smith,~R.~D. \emph{BMC bioinformatics} \textbf{2009}, \emph{10}, 87\relax
\mciteBstWouldAddEndPuncttrue
\mciteSetBstMidEndSepPunct{\mcitedefaultmidpunct}
{\mcitedefaultendpunct}{\mcitedefaultseppunct}\relax
\EndOfBibitem
\bibitem[Senko \latin{et~al.}(1995)Senko, Beu, and
  McLafferty]{senko1995determination}
Senko,~M.~W.; Beu,~S.~C.; McLafferty,~F.~W. \emph{Journal of the American
  Society for Mass Spectrometry} \textbf{1995}, \emph{6}, 229--233\relax
\mciteBstWouldAddEndPuncttrue
\mciteSetBstMidEndSepPunct{\mcitedefaultmidpunct}
{\mcitedefaultendpunct}{\mcitedefaultseppunct}\relax
\EndOfBibitem
\bibitem[O’Connor \latin{et~al.}(2006)O’Connor, Lin, Cournoyer, Pittman,
  Belyayev, and Budnik]{o2006long}
O’Connor,~P.~B.; Lin,~C.; Cournoyer,~J.~J.; Pittman,~J.~L.; Belyayev,~M.;
  Budnik,~B.~A. \emph{Journal of the American Society for Mass Spectrometry}
  \textbf{2006}, \emph{17}, 576--585\relax
\mciteBstWouldAddEndPuncttrue
\mciteSetBstMidEndSepPunct{\mcitedefaultmidpunct}
{\mcitedefaultendpunct}{\mcitedefaultseppunct}\relax
\EndOfBibitem
\bibitem[Tsybin \latin{et~al.}(2007)Tsybin, He, Emmett, Hendrickson, and
  Marshall]{tsybin2007ion}
Tsybin,~Y.~O.; He,~H.; Emmett,~M.~R.; Hendrickson,~C.~L.; Marshall,~A.~G.
  \emph{Analytical chemistry} \textbf{2007}, \emph{79}, 7596--7602\relax
\mciteBstWouldAddEndPuncttrue
\mciteSetBstMidEndSepPunct{\mcitedefaultmidpunct}
{\mcitedefaultendpunct}{\mcitedefaultseppunct}\relax
\EndOfBibitem
\bibitem[Lermyte \latin{et~al.}(2015)Lermyte, {\L}{\k{a}}cki, Valkenborg,
  Baggerman, Gambin, and Sobott]{lermyte2015understanding}
Lermyte,~F.; {\L}{\k{a}}cki,~M.~K.; Valkenborg,~D.; Baggerman,~G.; Gambin,~A.;
  Sobott,~F. \emph{International Journal of Mass Spectrometry} \textbf{2015},
  \emph{390}, 146--154\relax
\mciteBstWouldAddEndPuncttrue
\mciteSetBstMidEndSepPunct{\mcitedefaultmidpunct}
{\mcitedefaultendpunct}{\mcitedefaultseppunct}\relax
\EndOfBibitem
\bibitem[Roepstorff and Fohlman(1984)Roepstorff, and Fohlman]{RoepstorffScheme}
Roepstorff,~P.; Fohlman,~J. \emph{Biomed. Mass Spectrom.} \textbf{1984},
  \emph{11}, 601\relax
\mciteBstWouldAddEndPuncttrue
\mciteSetBstMidEndSepPunct{\mcitedefaultmidpunct}
{\mcitedefaultendpunct}{\mcitedefaultseppunct}\relax
\EndOfBibitem
\bibitem[{\L}\k{a}cki \latin{et~al.}(2017){\L}\k{a}cki, Startek, Valkenborg,
  and Gambin]{lacki2017isospec}
{\L}\k{a}cki,~M.~K.; Startek,~M.; Valkenborg,~D.; Gambin,~A. \emph{Analytical
  Chemistry} \textbf{2017}, \emph{89}, 3272--3277\relax
\mciteBstWouldAddEndPuncttrue
\mciteSetBstMidEndSepPunct{\mcitedefaultmidpunct}
{\mcitedefaultendpunct}{\mcitedefaultseppunct}\relax
\EndOfBibitem
\bibitem[Valkenborg \latin{et~al.}(2012)Valkenborg, Mertens, Lemi\`{e}re,
  Witters, and Burzykowski]{Valkenborg2012Isotopic}
Valkenborg,~D.; Mertens,~I.; Lemi\`{e}re,~F.; Witters,~E.; Burzykowski,~T.
  \emph{Mass Spectrom. Rev.} \textbf{2012}, \emph{31}, 96--109\relax
\mciteBstWouldAddEndPuncttrue
\mciteSetBstMidEndSepPunct{\mcitedefaultmidpunct}
{\mcitedefaultendpunct}{\mcitedefaultseppunct}\relax
\EndOfBibitem
\bibitem[Cormen(2009)]{cormen2009leiserson}
Cormen,~T. \emph{Leiserson C. Rivest R., Stein C. Introduction to
  Algorithms.-3rd}; MIT Press, 2009\relax
\mciteBstWouldAddEndPuncttrue
\mciteSetBstMidEndSepPunct{\mcitedefaultmidpunct}
{\mcitedefaultendpunct}{\mcitedefaultseppunct}\relax
\EndOfBibitem
\bibitem[James \latin{et~al.}(2013)James, Witten, Hastie, and
  Tibshirani]{james2013introduction}
James,~G.; Witten,~D.; Hastie,~T.; Tibshirani,~R. \emph{An introduction to
  statistical learning}; Springer, 2013; Vol. 112\relax
\mciteBstWouldAddEndPuncttrue
\mciteSetBstMidEndSepPunct{\mcitedefaultmidpunct}
{\mcitedefaultendpunct}{\mcitedefaultseppunct}\relax
\EndOfBibitem
\bibitem[Andersen \latin{et~al.}(2013)Andersen, Dahl, and
  Vandenberghe]{andersen2013cvxopt}
Andersen,~M.~S.; Dahl,~J.; Vandenberghe,~L. \emph{Available at cvxopt. org}
  \textbf{2013}, \emph{54}\relax
\mciteBstWouldAddEndPuncttrue
\mciteSetBstMidEndSepPunct{\mcitedefaultmidpunct}
{\mcitedefaultendpunct}{\mcitedefaultseppunct}\relax
\EndOfBibitem
\bibitem[Edmonds and Karp(1972)Edmonds, and Karp]{edmonds1972theoretical}
Edmonds,~J.; Karp,~R.~M. \emph{Journal of the ACM (JACM)} \textbf{1972},
  \emph{19}, 248--264\relax
\mciteBstWouldAddEndPuncttrue
\mciteSetBstMidEndSepPunct{\mcitedefaultmidpunct}
{\mcitedefaultendpunct}{\mcitedefaultseppunct}\relax
\EndOfBibitem
\bibitem[Hagberg \latin{et~al.}(2008)Hagberg, Schult, and
  Swart]{hagberg-2008-exploring}
Hagberg,~A.~A.; Schult,~D.~A.; Swart,~P.~J. Exploring network structure,
  dynamics, and function using {NetworkX}. Proceedings of the 7th Python in
  Science Conference (SciPy2008). Pasadena, CA USA, 2008; pp 11--15\relax
\mciteBstWouldAddEndPuncttrue
\mciteSetBstMidEndSepPunct{\mcitedefaultmidpunct}
{\mcitedefaultendpunct}{\mcitedefaultseppunct}\relax
\EndOfBibitem
\bibitem[Gillespie(1977)]{gillespie1977exact}
Gillespie,~D.~T. \emph{The journal of physical chemistry} \textbf{1977},
  \emph{81}, 2340--2361\relax
\mciteBstWouldAddEndPuncttrue
\mciteSetBstMidEndSepPunct{\mcitedefaultmidpunct}
{\mcitedefaultendpunct}{\mcitedefaultseppunct}\relax
\EndOfBibitem
\bibitem[McLuckey and Stephenson(1999)McLuckey, and
  Stephenson]{McLuckey1999-su}
McLuckey,~S.~A.; Stephenson,~J.~L. \emph{Mass Spectrom. Rev.} \textbf{1999},
  \emph{17}, 369--407\relax
\mciteBstWouldAddEndPuncttrue
\mciteSetBstMidEndSepPunct{\mcitedefaultmidpunct}
{\mcitedefaultendpunct}{\mcitedefaultseppunct}\relax
\EndOfBibitem
\bibitem[Lermyte \latin{et~al.}(2015)Lermyte, Verschueren, Brown, Williams,
  Valkenborg, and Sobott]{lermyte2015characterization}
Lermyte,~F.; Verschueren,~T.; Brown,~J.~M.; Williams,~J.~P.; Valkenborg,~D.;
  Sobott,~F. \emph{Methods} \textbf{2015}, \emph{89}, 22--29\relax
\mciteBstWouldAddEndPuncttrue
\mciteSetBstMidEndSepPunct{\mcitedefaultmidpunct}
{\mcitedefaultendpunct}{\mcitedefaultseppunct}\relax
\EndOfBibitem
\bibitem[Efron and Tibshirani(1994)Efron, and
  Tibshirani]{efron1994introduction}
Efron,~B.; Tibshirani,~R.~J. \emph{An introduction to the bootstrap}; CRC
  press, 1994\relax
\mciteBstWouldAddEndPuncttrue
\mciteSetBstMidEndSepPunct{\mcitedefaultmidpunct}
{\mcitedefaultendpunct}{\mcitedefaultseppunct}\relax
\EndOfBibitem
\bibitem[Wasserman(2013)]{wasserman2013all}
Wasserman,~L. \emph{All of statistics: a concise course in statistical
  inference}; Springer Science \& Business Media, 2013\relax
\mciteBstWouldAddEndPuncttrue
\mciteSetBstMidEndSepPunct{\mcitedefaultmidpunct}
{\mcitedefaultendpunct}{\mcitedefaultseppunct}\relax
\EndOfBibitem
\end{mcitethebibliography}

\end{document}